\documentclass[
reprint,
superscriptaddress,
nofootinbib,
amsmath,amssymb,
aps,
onecolumn
]{revtex4-2}
\usepackage{graphicx}
\usepackage{hyperref}
\usepackage{dcolumn}
\usepackage{bm}
\usepackage[mathlines]{lineno}
\usepackage{verbatim}
\usepackage{hyperref}
\hypersetup{colorlinks,linkcolor={blue},citecolor={blue},urlcolor={blue}}
\begin{document}
\preprint{APS/123-QED}
\title{Effect of Spin-Dependent Short-Range Correlations on Nuclear Matrix Elements for Neutrinoless Double Beta Decay of $^{48}$Ca}
\author{S.Sarkar}
\email{shahariar.ph@sric.iitr.ac.in}
\affiliation{Department of Physics, Indian Institute of Technology Roorkee, Roorkee - 247667, Uttarakhand, India}
\author{Y. Iwata}
\email{iwata\_phys@08.alumni.u-tokyo.ac.jp}
\affiliation{Osaka University of Economics and Law, Yao, Osaka 581-0853, Japan}
\date{\today}
\begin{abstract}
The neutrinoless double beta decay is a pivotal weak nuclear process that holds the potential to unveil the Majorana nature of neutrinos and predict their absolute masses. In this study, we delve into examining the impact of spin-dependent short-range correlations (SRC) on the nuclear matrix elements (NMEs) for the light neutrino-exchange mechanism in neutrinoless double beta ($0\nu\beta\beta$) decay of $^{48}$Ca, employing an extensive interacting nuclear shell model.
All computations are performed employing the effective shell model Hamiltonian GXPF1A, encompassing the entire $fp$ model space through the closure approximation. Our investigation examines the NMEs' dependencies on factors such as the number of intermediate states, coupled spin-parity attributes of neutrons and protons, neutrino momentum, inter-nucleon separation, and closure energy. This scrutiny is performed with respect to both the conventional Jastrow-type approach of SRC, employing various parameterizations, and the spin-dependent SRC paradigm.
Our findings illuminate a discernible distinction in NMEs induced by spin-dependent SRC, differing by approximately 10-20\% from those computed through the conventional Jastrow-type SRC, incorporating distinct parameterizations.
\end{abstract}

\maketitle

\section{Introduction}

The neutrinoless double beta decay ($0\nu\beta\beta$) is a rare and crucial weak nuclear decay that occurs in certain even-even nuclei such as $^{48}$Ca, $^{76}$Ge, $^{82}$Se, $^{96}$Zr, $^{100}$Mo, $^{116}$Cd, $^{124}$Sn, $^{130}$Te, $^{126}$Xe, etc. In this process, two neutrons inside the decaying nucleus are converted into two protons and two electrons without emitting any neutrinos that violate the lepton number conservation. The neutrino appears as a virtual intermediate Majorana particle \cite{Dolinski:2019nrj,Vergados:2016hso,engel2017status,RevModPhys.80.481}. Although predicted long back in 1939 by Wolfgang Furry \cite{PhysRev.56.1184,vergados2012theory} based on E.Majorana's symmetric theory for fermion and anti-fermion \cite{majorana1937teoria} followed by G. Racah's chain reactions \cite{racah1937sulla} in 1937, this rare process is not observed yet in the experiment. But, observing this rare process would prove that neutrinos are Majorana particles, which is indeed widely favored in many beyond the standard model (BSM) physics theories for the explanation of the smallness of neutrino mass \cite{deppisch2012neutrinoless,PhysRevD.25.2951,rodejohann2011neutrino}. Also, there is no definite value of absolute neutrino mass measured yet in the experiment. An upper limit for absolute neutrino mass of 0.8 eV has been derived through the tritium beta decay experiment-KATRIN \cite{KATRIN:2021uub} and the precise upper limit on the effective Majorana neutrino mass of 36–156 meV is predicted in $0\nu\beta\beta$ decay experiment KamLAND-Zen on $^{136}$Xe \cite{KamLAND-Zen:2022tow}. In this context, the $0\nu\beta\beta$ process can also provide information on the absolute mass scale of neutrinos \cite{tomoda1991double, RevModPhys.80.481} which makes this process more interesting. 

A number of particle physics mechanisms have been proposed for the $0\nu\beta\beta$ decay process. One of them is the widely studied standard light neutrino-exchange mechanism \cite{rodin2006assessment,PhysRevC.60.055502}, which will be the focus of this study as well. Other mechanisms include the heavy neutrino-exchange mechanism \cite{vergados2012theory}, the left-right symmetric mechanisms \cite{PhysRevLett.44.912,PhysRevLett.47.1713}, and the supersymmetric particle exchange mechanisms \cite{PhysRevD.34.3457,vergados1987neutrinoless} etc.  

The half-life for $0\nu\beta\beta$ decay is connected with absolute neutrino mass through a quantity called nuclear matrix elements (NMEs), which are theoretically calculated in different nuclear many-body models \cite{engel2017status} such as quasiparticle random phase approximation (QRPA) \cite{vergados2012theory}, the interacting shell-model (ISM) \cite{PhysRevLett.100.052503,PhysRevC.81.024321,PhysRevC.88.064312,PhysRevLett.113.262501,PhysRevLett.116.112502}, the interacting boson model (IBM) \cite{PhysRevC.79.044301,PhysRevLett.109.042501}, the generator coordinate method (GCM) \cite{PhysRevLett.105.252503}, the energy density functional (EDF) theory \cite{PhysRevLett.105.252503,PhysRevC.90.054309}, the relativistic energy density functional (REDF) theory \cite{PhysRevC.90.054309,PhysRevC.91.024316}, and the projected Hartree-Fock Bogolibov model (PHFB) \cite{PhysRevC.82.064310}, $ab$ $initio$ variational Monte Carlo (VMC) technique \cite{PhysRevC.97.014606,wang2019comparison,PhysRevC.100.055504} etc. The accuracy of NMEs plays an important role in predicting the correct value for the half-life of $0\nu\beta\beta$ decay and absolute neutrino mass. Hence, calculating the NMEs accurately is one of the central themes of research for studying the $0\nu\beta\beta$ decay. 

All over the globe, there are several Major experimental investigations are being carried out for different $0\nu\beta\beta$ decaying candidates such as GERDA($^{76}$Ge), MAJORANA DEMONSTRATOR($^{76}$Ge), LEGEND($^{76}$Ge), CUORE($^{130}$Te), CUPID($^{82}$Se/$^{100}$Mo/$^{130}$Te), AMoRE($^{100}$Mo), EXO-200($^{136}$Xe), nEXO($^{136}$Xe), NEXT($^{136}$Xe), PandaX-III($^{136}$Xe), KamLAND-Zen and KamLAND2-Zen($^{136}$Xe), CANDLES( $^{48}$Ca), Super-NEMO($^{82}$Se) \cite{Dolinski:2019nrj}. Our particular interest in this study is 0$\nu\beta\beta$ decay of $^{48}$Ca which is one of the simplest decay candidates with significant experimental interest. Particularly, we aim to examine the effect of spin-dependent SRC on the NMEs for the 0$\nu\beta\beta$ decay of $^{48}$Ca using the nuclear shell model. 

The occurrence of the $0\nu\beta\beta$ decay in $^{48}$Ca unfolds as:
\begin{equation}
    ^{48}\text{Ca}\rightarrow^{48}\text{Ti}+e^-+e^-.
\end{equation}
In previous studies, the computation of NMEs corresponding to the light neutrino-exchange mechanism of the $0\nu\beta\beta$ decay in $^{48}$Ca was carried out (Refs. \cite{PhysRevC.102.034317,PhysRevC.101.014307,PhysRevC.101.035504,PhysRevC.98.035502,PhysRevLett.113.262501,PhysRevC.88.064312,PhysRevC.87.014320,PhysRevC.86.067304,PhysRevC.81.024321,Menendez:2008jp}). In these investigations, the influences of SRC were integrated through a Jastrow-type approach \cite{PhysRevC.79.055501,vogel2012nuclear}, characterizing the radial-dependence effects of SRC. An alternative method, known as the Unitary Correlation Operator Method (UCOM) \cite{feldmeier1998unitary,NEFF2003311,ROTH20043}, also accounted for SRC via radial dependence. More recently, in a notable advancement presented in Ref. \cite{PhysRevC.90.065504}, the interplay of spin in conjunction with SRC was incorporated to compute NMEs for the light neutrino-exchange $0\nu\beta\beta$ decay in $^{48}$Ca. This study, however, operated within a confined valence space, focusing exclusively on the $f_{7/2}$ orbital within the $pf$-shell.

Motivated by these strides, our current endeavor aims to expand the computational scope. Specifically, we engage the complete $fp$ model space encompassing the orbitals $0f_{7/2}$, $0f_{5/2}$, $1p_{3/2}$, and $1p_{1/2}$ for $^{48}$Ca. This comprehensive approach enables us to thoroughly assess the cumulative effect of spin on SRC for the $0\nu\beta\beta$ decay in $^{48}$Ca. Furthermore, we undertake an in-depth exploration of the variances between the spin-dependent SRC scenario and the conventional Jastrow-type SRC approach. We systematically investigate the dependencies of NMEs on factors such as the coupled spin-parity of protons and neutrons, the number of intermediate states, neutrino momentum, inter-nucleon separation, and closure energy. Our goal is to uncover how these dependencies diverge under the influence of spin-dependent SRC and to contrast them with the outcomes of the traditional Jastrow-type SRC approach. As our investigation progresses, we envision extending this approach to encompass other decay candidates within the $0\nu\beta\beta$ realm, thereby offering a more comprehensive understanding of this intriguing nuclear process.

Driven by this motivation, the structure of this paper unfolds as follows. Section \ref{sec:II} outlines the formulation of the decay rate, NMEs, and transition operators pertinent to the $0\nu\beta\beta$ decay. In Section \ref{sec:III}, we delve into the integration of SRC effects within both the Jastrow-type approach and the spin-dependent approach. The methodology for calculating NMEs using the shell model within the closure method is detailed in Section \ref{sec:IV}. Subsequently, Section \ref{sec:V} presents the comprehensive results and initiates discussions around the observed outcomes. Lastly, Section \ref{sec:VI} encapsulates the synthesis of our findings and delineates potential avenues for future exploration.

\section{\label{sec:II}The decay rate and NMEs for $0\nu\beta\beta$ decay}
The decay rate for the light neutrino-exchange mechanism of $0\nu\beta\beta$ decay can be written as \cite{PhysRevC.60.055502}
\begin{equation}
    [T^{0\nu}_\frac{1}{2}]^{-1}=G^{0\nu}|M^{0\nu}|^2\left(\frac{\langle m_{\beta\beta}\rangle}{m_e}\right)^2,
\end{equation}
where $G^{0\nu}$ is the phase-space factor that can be calculated accurately \cite{PhysRevC.85.034316}, $M^{0\nu}$ is the total nuclear matrix element for the light neutrino-exchange mechanism, and $m_{\beta\beta}$ is the effective Majorana neutrino mass defined by the neutrino mass eigenvalues $m_k$ and the neutrino mixing matrix elements $U_{ek}$
\begin{equation}
\langle m_{\beta\beta}\rangle= \left\lvert\sum_k m_k U_{ek}^2\right\rvert.
\end{equation}
The total nuclear matrix element $M^{0\nu}$ is the sum of Gamow-Teller ($M_{GT}$), Fermi ($M_{F}$), and tensor ($M_{T}$) matrix elements, as given by \cite{RevModPhys.80.481}:
\begin{equation}
M^{0\nu}=M_{GT}-\left(\frac{g_V}{g_A}\right)^{2}M_{F}+M_{T},
\end{equation}
where $g_V$ and $g_A$ are the vector and axial-vector constants, respectively. In this study, $g_V$=1 and the bare value of $g_A$=1.27 is used. The matrix elements $M_{GT}$, $M_{F}$, and $M_{T}$ of transition operator $O_{12}^\alpha$ of $0\nu\beta\beta$ decay are expressed as \cite{PhysRevLett.113.262501}:
\begin{eqnarray}
\label{Eq:NMEMAIN}
&&M_{\alpha}=\langle f|\mathcal{O}_{12}^\alpha|i\rangle
\end{eqnarray}
where $\alpha\in{F, GT, T}$, $|i\rangle$ corresponds to the $0^+$ ground state of the parent nucleus $^{48}$Ca in the present study, and $|f\rangle$ corresponds to the $0^+$ ground state of the granddaughter nucleus $^{48}$Ti. 

The calculation of two-body matrix elements (TBMEs) for $0\nu\beta\beta$ decay scalar two-particle transition operators $O_{12}^{\alpha}$ have both spin and radial neutrino potential parts. These operators are given by \cite{PhysRevC.88.064312}:
\begin{eqnarray}
\label{eq:ncoperator}
O_{12}^{GT}&&=\tau_{1-}\tau_{2-}(\boldsymbol{\sigma_1.\sigma_2})H_{GT}(r,E_k),
\nonumber\\
O_{12}^{F}&&=\tau_{1-}\tau_{2-}H_{F}(r,E_k),
\\
O_{12}^{T}&&=\tau_{1-}\tau_{2-}S_{12}H_{T}(r,E_k),
\nonumber
\end{eqnarray}
where, $\tau$ is isospin annihilation operator, $\boldsymbol{r=r_1-r_2}$ is the inter-nucleon distance of the decaying nucleons, and $r=\boldsymbol{|r|}$. The operator $S_{12}$ is defined as $S_{12}=3(\boldsymbol{\sigma_1 .\hat{r})(\sigma_2.\hat{r})-(\sigma_1.\sigma_2)}$.
For the light-neutrino exchange mechanism of $0\nu\beta\beta$ decay, the radial neutrino potential with explicit dependence on the energy of the intermediate states is given by \cite{PhysRevC.88.064312}:
\begin{equation}
\label{eq:npnc}
H_\alpha (r,E_{k})=\frac{2R}{\pi}\int_{0}^{\infty}\frac{f_\alpha(q,r)qdq}{q+E_{k}-(E_{i}+E_{f})/2}, 
\end{equation}
where $R$ is the radius of the parent nucleus, $q$ is the momentum of the virtual Majorana neutrino, $E_{i}$, $E_{k}$ and $E_{f}$ is the energy of initial, intermediate, and final nuclei, and $f_\alpha(q,r)=j_p(q,r)h_\alpha(q,r)$ with $j_p(q,r)$ is spherical Bessel function with $p=0$ for Fermi and Gamow-Teller type NMEs and $h_\alpha(q,r)$ is the term that accounts for the effects of higher-order currents (HOC) and finite nucleon size (FNS) \cite{PhysRevC.60.055502} and given by \cite{PhysRevC.60.055502,PhysRevC.79.055501}:
\begin{eqnarray}
   h_F(q^2)=&&g_{V}^{2}(q^2),\\
   h_{GT}(q^2)=&&\frac{g_{A}^{2}(q^2)}{g_{A}^2}\left(1-\frac{2}{3}\frac{q^2}{q^2+m_\pi^2}+\frac{1}{3}\left(\frac{q^2}{q^2+m_\pi^2}\right)^2\right)\nonumber\\
    &&+\frac{2}{3}\frac{g_M^2(q^2)}{g_A^2}\frac{q^2}{4m_p^2},\\
   h_{T}(q^2)=&&\frac{g_{A}(q^{2})}{g_{A}^{2}}\left(\frac{2}{3}\frac{q^{2}}{q^{2}+m_{\pi}^{2}}-\frac{1}{3}\left(\frac{q^{2}}{q^{2}+m_{\pi}^{2}}\right)^{2}\right)\nonumber\\
   &&+\frac{1}{3}\frac{g_{M}^{2}(q^{2})}{g_{A}^{2}}\frac{q^{2}}{4m_{p}^{2}}.
   \end{eqnarray}
In this regard, the $g_V(q^2)$, $g_A(q^2)$, and $g_M(q^2)$ form factors, which account for FNS effects, are used. In the dipole approximation, the form factors are given by the following equations \cite{PhysRevC.60.055502,PhysRevC.81.024321}:
 \begin{eqnarray}
        g_V(q^2)=&&\frac{g_V}{\left(1+  \frac{q^2}{M_V^2}\right)^2},\\
     g_A(q^2)=&&\frac{g_A}{\left(1+  \frac{q^2}{M_A^2}\right)^2},\\
     g_M(q^2)=&&(\mu_p-\mu_n)g_V(q^2).
 \end{eqnarray}
where $g_V$, $g_A$, $\mu_p$, and $\mu_n$ are the vector, axial-vector, and magnetic moment coupling constants for the nucleon, and $M_V$, and $M_A$ are the vector and axial-vector meson masses, respectively. The values of $M_V$ and $M_A$ are 850 MeV and 1086 MeV, respectively, while $\mu_p-\mu_n$ is 4.7 is used in the calculation \cite{PhysRevC.81.024321}. The masses of the proton and pion are denoted by $m_p$ and $m_\pi$, respectively.
\section{\label{sec:III}The effects of short-range correlations on $0\nu\beta\beta$ decay}
In the computation of the NMEs for $0\nu\beta\beta$ decay, it becomes imperative to factor in the influences of SRC, as discussed previously. Over time, various techniques have been employed to integrate SRC effects into the NME calculations for $0\nu\beta\beta$ decay. One of the extensively utilized and classical methodologies is known as the Jastrow approach, as discussed in references \cite{PhysRevC.79.055501} and \cite{vogel2012nuclear}. Moreover, a novel spin-dependent approach has emerged, as outlined in reference \cite{PhysRevC.90.065504}. Both of these methodologies are delineated comprehensively below.

\subsection{\textbf{The Jastrow approach}}
A standard method to include SRC is via a phenomenological Jastrow-like function \cite{PhysRevC.79.055501,vogel2012nuclear}. By including the SRC effect in the Jastrow approach, one can write the NMEs of $0\nu\beta\beta$ defined in Eq. (\ref{Eq:NMEMAIN}) as
 \cite{PhysRevC.79.055501}
\begin{eqnarray}
\label{eq:srcmain}
&&{M}_{\alpha}=\langle f|f_{Jastrow}(r)O_{12}^\alpha f_{Jastrow}(r)|i\rangle,
\end{eqnarray}
where the Jastrow-type SRC function is defined as 
\begin{equation}
\label{eq:src}
  f_{Jastrow}(r)=1-ce^{-ar^{2}}(1-br^{2}).  
\end{equation}
In literature, three different SRC parametrizations are used: Miller-Spencer, Charge-Dependent Bonn (CD-Bonn), and Argonne V18 (AV18) to parametrize $a, b,$ and $c$ \cite{PhysRevC.81.024321}. The parameters $a$, $b$, and $c$ in different SRC parametrizations are given in Table~\ref{tab:src}. This approach of using a Jastrow-like function to include the effects of SRC is extensively used in Refs.  \cite{PhysRevC.81.024321,neacsu2012fast,Menendez:2008jp}.

\begin{table}[h!]
\caption{Parameters for the Jastrow-type function of Eq. (\ref{eq:src})\label{tab:src}}.
\begin{ruledtabular}
\begin{tabular}{cccc}
\textbf{SRC Type} & \textbf{a} & \textbf{b} & \textbf{c}\\
\hline
Miller-Spencer&1.10&0.68&1.00\\
CD-Bonn&1.52&1.88&0.46\\
AV18&1.59&1.45&0.92
\end{tabular}
\end{ruledtabular}
\end{table}

\subsection{\textbf{Spin-Dependent approach}}
Recently in Ref. \cite{PhysRevC.90.065504}, for the first time short nature of nucleon-nucleon correlations was taken care of based on the dependence of the SRC operator in spin and coordinate space. 
The new spin-dependent SRC correlation function is defined as
\begin{equation}
\label{eq:src-spin-dependent}
    f_{SD}(r)=f(r)+g(r)\boldsymbol{\sigma_{1}.\sigma_{2}}. 
\end{equation}
The transition operators for $0\nu\beta\beta$ decay by incorporating spin-dependent SRC are written as
\begin{equation}
\widetilde{O}_{12}^{\alpha}=f_{SD}(r)O_{12}^{\alpha}f_{SD}(r).  
\end{equation}
Using the relation $\boldsymbol{(\sigma_{1}.\sigma_{2})^{2}=(3-2(\sigma_{1}.\sigma_{2}))}$, the new Fermi operator of $0\nu\beta\beta$ decay is written as
\begin{equation}
    \widetilde{O}_{12}^{F}=[f(r)^{2}+3g(r)^{2}]H_{F}(r)+(\boldsymbol{\sigma_{1}.\sigma_{2}})2g(r)[f(r)-g(r)]*H_{F}(r)
\end{equation}
and the Gamow-Teller operator of $0\nu\beta\beta$ decay are written as
\begin{equation}
\widetilde{O}_{12}^{GT}=[f(r)^{2}-4g(r)f(r)+7g(r)^{2}]H_{GT}(r)(\boldsymbol{\sigma_{1}.\sigma_{2}})+6g(r)[f(r)-g(r)]H_{GT}(r)
\end{equation}
The function $f(r)$ and $g(r)$ are defined as \cite{PhysRevC.90.065504,ecorrelation,valli2007shear}
\begin{equation}
f(r)=a_{1}-b_{1}e^{-c_{1}r^{2}}+d_{1}e^{-e_{1}(r-f_{1})^{2}},
\end{equation}

\begin{equation}
g(r)=a_{2}e^{-b_{2}r^{2}}(1+c_{2}r+d_{2}r^{2}).
\end{equation}
The different parameters of $f(r)$ and $g(r)$ are presented in Table \ref{tab:spin-src}. Explicit details of using this SRC approach can be found in Ref. \cite{ecorrelation}. Derivation of the above formalism can be found in Ref. \cite{valli2007shear}.

\begin{table}[h!]
\caption{Parameters of $f(r)$ and $g(r)$ for the spin-dependent SRC\label{tab:spin-src}}.
\begin{ruledtabular}
\begin{tabular}{cccc}
\textbf{Parameters for $f(r)$} & \textbf{Value} & \textbf{Parameters for $g(r)$} & \textbf{Value}\\
\hline
$a_{1}$&1.00&$a_{2}$&0.04\\
$b_{1}$&0.92&$b_{2}$&1.39\\
$c_{1}$&2.56&$c_{2}$&2.92\\
$d_{1}$&0.33&$d_{2}$&-5.97\\
$e_{1}$&0.57\\
$f_{1}$&-0.94\\
\end{tabular}
\end{ruledtabular}
\end{table}

In addition to the above, the authors of Refs. \cite{kortelainen2007short,PhysRevC.75.051303} have recently proposed another method namely: the Unitary Correlation Operator Method \cite{feldmeier1998unitary,NEFF2003311,ROTH20043} to estimate the effects of SRC, which leads to a much smoother correction \cite{vogel2012nuclear}. In the present study, we only focus on using a standard Jastrow-type approach and spin-dependent approach to estimate the effects of SRC. Detailed descriptions of incorporating the SRC effects in different approaches can be found in Refs. \cite{vogel2012nuclear,vsimkovic20090}
\section{\label{sec:IV}The closure method of NMEs calculation for $0\nu\beta\beta$ decay through $(n-2)$ channel}
Now we discuss, including all the effects, how NMEs are calculated in the nuclear shell model. First of all, there are two approaches; one is nonclosure and the other is the closure approach. The nonclosure approach \cite{PhysRevC.88.064312,PhysRevC.102.034317,PhysRevC.101.014307} considers the effects of excitation energies of a large number of intermediate states of the true virtual nucleus ($^{21}$Sc in the present case) explicitly through the denominator of Eq. (\ref{eq:npnc}). In closure approximation \cite{PhysRevC.81.024321}, one approximates the term $E_{k} - (E_{i}+E_{f})/2$ in the denominator of the neutrino potential of Eq. (\ref{eq:npnc}) with a constant closure energy ($\langle E\rangle$) value such that 
\begin{equation}
    [E_{k}-(E_{i}+E_{f})/2]\rightarrow \langle E\rangle. 
\end{equation}Closure approximation avoids the complexity of calculating a large number of intermediate states, which can be computationally challenging for nuclear shell models for higher mass isotopes. The difficult part of closure approximation is picking the correct closure energy which has no definite method yet and can greatly influence the accuracy of NMEs. Recently, there are many studies of nonclosure approach for $0\nu\beta\beta$ decay of $^{48}$Ca \cite{PhysRevC.88.064312,PhysRevC.102.034317,PhysRevC.101.014307}, which predicted that $\langle E\rangle$=0.5 MeV is an optimal value of closure energy which gives NME in closure method very close to NME in nonclosure method. Hence, we use the closure method of NMEs calculation with closure energy value $\langle E\rangle$=0.5 MeV. 

Based on closure approach the $M_{GT}$, $M_{F}$ and $M_{T}$ matrix elements of the scalar two-body transition operator $O_{12}^\alpha$  of $0\nu\beta\beta$
can be expressed as the sum over the product of the two-body transition density (TBTD) and anti-symmetric two-body matrix elements $(\langle k_1',k_2',JT|\tau_{-1}\tau_{-2}O_{12}^\alpha|k_1,k_2,JT\rangle_A)$. For this first, we write the partial NME as a function of coupled spin-parity ($J^{\pi}$) of protons and neutrons and number of intermediate states ($N_m$) of TBTD calculation as \cite{PhysRevLett.113.262501}
\begin{eqnarray}
\label{eq:nmepartial}
&&M_{\alpha}(J^{\pi},N_m)=\sum_{k_1'\leq k_2',k_1\leq k_2}\nonumber\\
&&\mathrm{TBTD}(f,m,i,J^\pi)\langle k_1',k_2',J^\pi T|\tau_{-1}\tau_{-2}O_{12}^\alpha|k_1,k_2,J^\pi T\rangle_A,\nonumber\\
\end{eqnarray}
such that, the final required Fermi, GT, and tensor type NME of Eq. (\ref{Eq:NMEMAIN}) can be written as 
\begin{eqnarray}
{M}_{\alpha}=\sum_{J,N_{m}\leqslant N_c}{M}_{\alpha}(J^{\pi},N_{m}), 
\end{eqnarray}
where, $\alpha={(F,GT,T)}$, $J^{\pi}$ is the coupled spin-parity of two decaying neutrons or two final created protons, $\tau_{-}$ is the isospin annihilation operator, $A$ denotes that the two-body matrix elements are obtained using anti-symmetric two-nucleon wavefunctions, and $k$ stands for the set of spherical quantum numbers $(n; l; j)$, $N_c$ is cut-off on number of states of intermediate nucleus ($^{46}$Ca in the present case) for TBTD calculations. In our case, $|i\rangle$ is $0^+$ g.s. of the parent nucleus $^{48}$Ca,  $|f\rangle$ is the $0^+$ g.s. of the granddaughter nucleus $^{48}$Ti, and $k$ has the spherical quantum numbers for $0f_{7/2}$, $0f_{5/2}$, $1p_{3/2}$, and $1p_{1/2}$ orbitals. The TBTD can be expressed as \cite{PhysRevLett.113.262501}
\begin{eqnarray}
\mathrm{TBTD}(f,m,i,J)
=\langle f||[A^+(k_1', k_2' ,J)\otimes\tilde{A}(k_1, k_2 ,J) ]^{(0)}||i\rangle,
\end{eqnarray}
where,  
\begin{equation}
  A^+(k_1', k_2' ,J)=\frac{[ a^{+}(k_1')\otimes a^{+} (k_2')]^{J}_{M}}{\sqrt{1+\delta_{{k_1'}{k_2'}}}},
\end{equation} and 
\begin{equation}
   \tilde{A}(k_1, k_2 ,J)=(-1)^{J-M}A^+(k_1, k_2,J,-M)
\end{equation}
are the two-particle creation and annihilation operators of rank $J$, respectively. 

To evaluate TBTD, one needs a large number of two-nucleon transfer amplitudes (TNA). The TNA are calculated with a large set of intermediate states $|m\rangle$ of the (n-2) nucleon system ($^{46}$Ca in the present study), where n is the number of nucleons for the parent nucleus. The TBTD in terms of TNA is expressed as \cite{PhysRevLett.113.262501}
\begin{eqnarray}
\label{Eq:tna*tna}
\mathrm{TBTD}(f,m,i,J)=\sum_{m}\mathrm{TNA}(f,m,k_1', k_2', J_m)\mathrm{TNA}(i,m,k_1, k_2, J_m),\nonumber\\
\end{eqnarray}
where TNA are given by
\begin{equation}
  \mathrm{TNA}(f,m,k_1', k_2', J_m)=\frac{\langle f||A^+(k_1', k_2' ,J)||m\rangle}{\sqrt{2J_0+1}}.
\end{equation}
Here, $J_m$ is the spin of the allowed states of $^{46}$Ca. $J_0$ is spin of $|i\rangle$ and $|f\rangle$. $J_m$=$J$ when $J_0$=0 \cite{PhysRevLett.113.262501}.

\section{\label{sec:V}Results and discussion}
\begin{table*}
\caption{NMEs for $0\nu\beta\beta$ decay (the light neutrino-exchange mechanism) of $^{48}$Ca calculated with nuclear shell model for spin-dependent SRC and Jastrow-type SRC with different parametrization. Calculations are conducted within the closure approximation in the (n-2) channel, employing the GXPF1A effective shell model Hamiltonian. \label{tab:nmecomparison}}
\begin{ruledtabular}
\begin{tabular}{ccc}
NME Type&SRC Type&NME Value\\ \hline
$M_{F}$&None&-0.215\\
$M_{F}$&Miller-Spencer&-0.144\\
$M_{F}$&CD-Bonn&-0.232\\
$M_{F}$&AV18&-0.213\\
$M_{F}$&Spin-Dependent SRC&-0.190\\
\\
$M_{GT}$&None&0.774\\
$M_{GT}$&Miller-Spencer&0.540\\
$M_{GT}$&CD-Bonn&0.806\\
$M_{GT}$&AV18&0.740\\
$M_{GT}$&Spin-Dependent SRC&0.674\\
\\
$M^{0\nu}$&None&0.873\\
$M^{0\nu}$&Miller-Spencer&0.629\\
$M^{0\nu}$&CD-Bonn&0.950\\
$M^{0\nu}$&AV18&0.872\\
$M^{0\nu}$&Spin-Dependent SRC&0.792\\
\end{tabular}
\end{ruledtabular}
\end{table*}
The TNA are evaluated utilizing the nuclear shell model. For this purpose, we employ the KSHELL code \cite{Shimizu:2019xcd}, which facilitates the computation of essential wave functions and energies concerning the initial, intermediate, and final nuclei involved in the $0\nu\beta\beta$ decay process of $^{48}$Ca. Subsequently, these calculated wave functions are harnessed to determine the TNA, a pivotal component within the expression of NMEs for the $0\nu\beta\beta$ decay. 

For these calculations, we utilize the shell model Hamiltonian GXPF1A \cite{PhysRevC.82.064304}, tailored for the $fp$ model space, serving as a foundational input for our computations.

The $fp$ model space comprises the $f_{7/2}$, $p_{3/2}$, $f_{5/2}$, and $p_{1/2}$ orbitals. The computation of the two-body matrix element (TBME) components within the NMEs is executed using a proprietary code developed by us. The outcomes of distinct NME types, computed through the shell model approach with the Jastrow methodology employing various SRC parametrizations, and spin-dependent SRC are presented in Table \ref{tab:nmecomparison}.

It's important to note that the presentation of tensor-type NMEs is omitted from Table \ref{tab:nmecomparison} due to the lack of a solvable expression for the TBMEs inclusive of the spin-dependent SRC function specified in Eq. (\ref{eq:src-spin-dependent}). This is not a critical concern as the Gamow-Teller (GT) and Fermi-type NMEs overwhelmingly dominate over tensor-type NMEs. Therefore, in Table \ref{tab:nmecomparison}, the calculation of total NMEs solely considers the Fermi and GT-type contributions.

The NMEs corresponding to the application of spin-dependent SRC are approximately 10\% diminished in comparison to those calculated with AV18-type NMEs, as well as SRC-none scenarios, across Fermi, Gamow-Teller (GT), and total NMEs. In contrast, when gauged against CD-Bonn-type, the NMEs with spin-dependent SRC experience a more pronounced reduction, approximately 20\%, across Fermi, GT, and total NMEs.

Remarkably, among the various SRC parametrizations, the Miller-Spencer type NMEs manifest the most pronounced impact of SRC. Notably, NMEs with spin-dependent SRC showcase an expansion of approximately 20\% relative to Miller-Spencer type NMEs, encompassing the realms of Fermi, GT, and total NMEs.
\subsection{\textbf{Dependence of NMEs on coupled spin-parity ($J^\pi$) of two decaying neutrons and two created protons}}

Now we examine the comparison of contribution from different coupled spin-parity in the NMEs for CD-Bonn type NME and spin-dependent SRC.
To do this, we decompose NMEs in terms of partial nuclear matrix elements of coupled spin-parity ($J^\pi$) of two decaying neutrons or two created protons as 
\begin{equation}
M_{\alpha}(J^{\pi})=\sum_{N_{m}\leq N_{c}}M_{\alpha}(J^\pi,N_{m}),
\end{equation}
where, one can define $M^{0\nu}_{\alpha}(J^\pi)$ using Eq.~(\ref{eq:nmepartial}). It should be noted that $j^{\pi}$ in the above equation also represents the spin-parity of the states of intermediate nucleus $^{46}$Ca.

\begin{figure}
\includegraphics[trim=0cm 4cm 0cm 2cm,width=\linewidth]{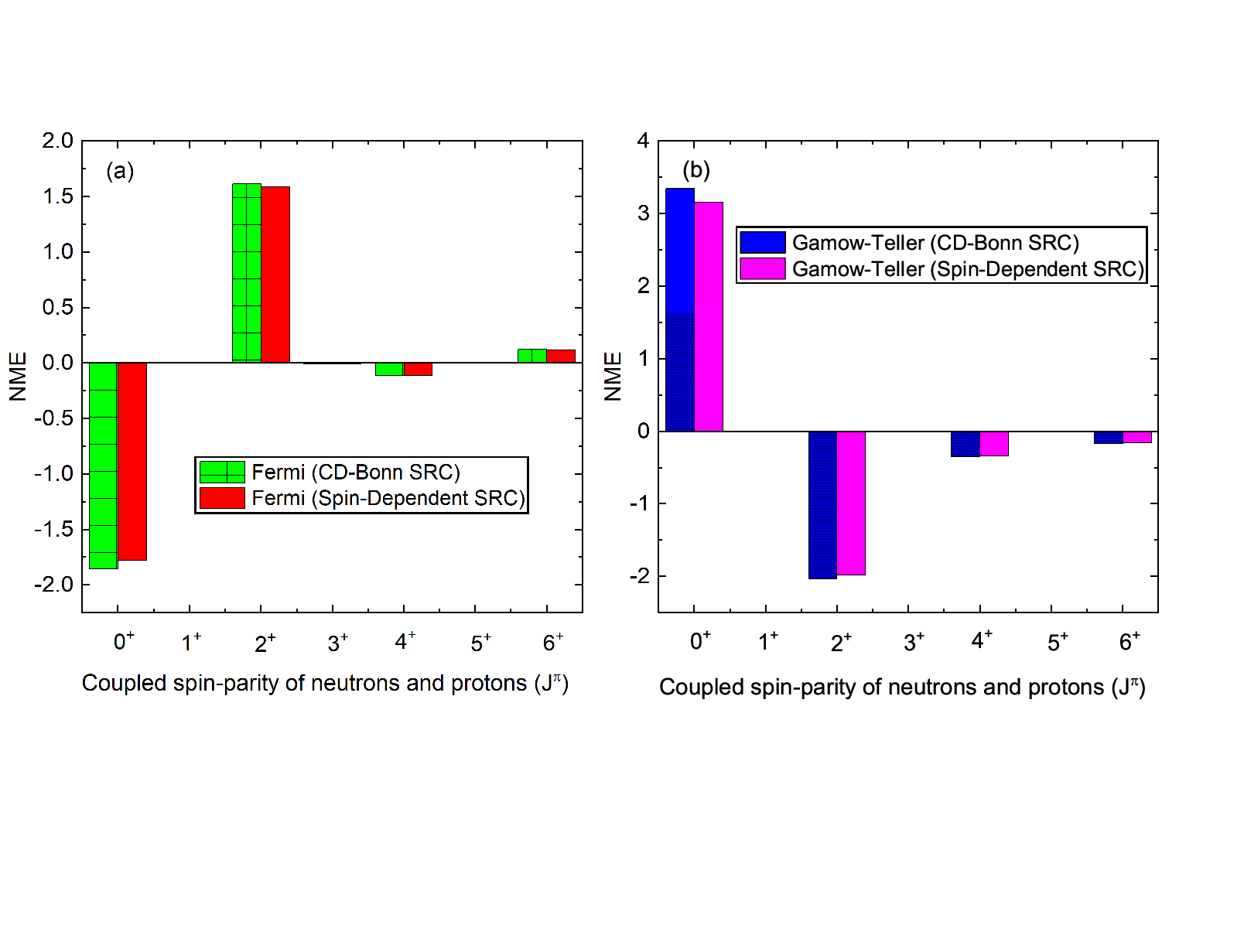}
\caption{\label{fig:NMEvsJ}(Color online) Contribution of various coupled spin-parity states ($J^{\pi}$) of two initial neutrons or two final created protons for (a) Fermi-type and (b) Gamow-Teller-type NMEs. The illustrated NMEs pertain to the light neutrino-exchange mechanism in the context of $0\nu\beta\beta$ decay of $^{48}$Ca. Calculations have been carried out utilizing the nuclear shell model, considering both CD-Bonn and spin-dependent SRC.
}
\end{figure}

The contributions of NMEs through different $J^\pi$ are shown in Fig.~\ref{fig:NMEvsJ}.
We observe that for all types of NMEs, the most dominant contributions come from the 0$^{+}$ and 2$^{+}$ states. Additionally, the contribution from 0$^{+}$ and 2$^{+}$ states have opposite signs, leading to a reduction in the total NMEs. There are also small contributions from the 4$^{+}$ and 6$^{+}$ states, with almost negligible contributions from odd-$J^\pi$ states. This is due to the pairing effect, which is responsible for the dominance of even-$J^\pi$ contributions \cite{PhysRevLett.113.262501}. Note that spin-dependent SRCs are smaller in magnitude than CD-Bonn type SRC parametrizations. But, the overall pattern of dependence on $J^{\pi}$ is very similar for both SRCs.  
\subsection{\textbf{Variation of NMEs for $0\nu\beta\beta$ with the cutoff number of states ($N_c$) of $^{46}$Ca}}
\begin{figure}
\includegraphics[trim=0cm 6cm 0cm 2cm,width=\linewidth]{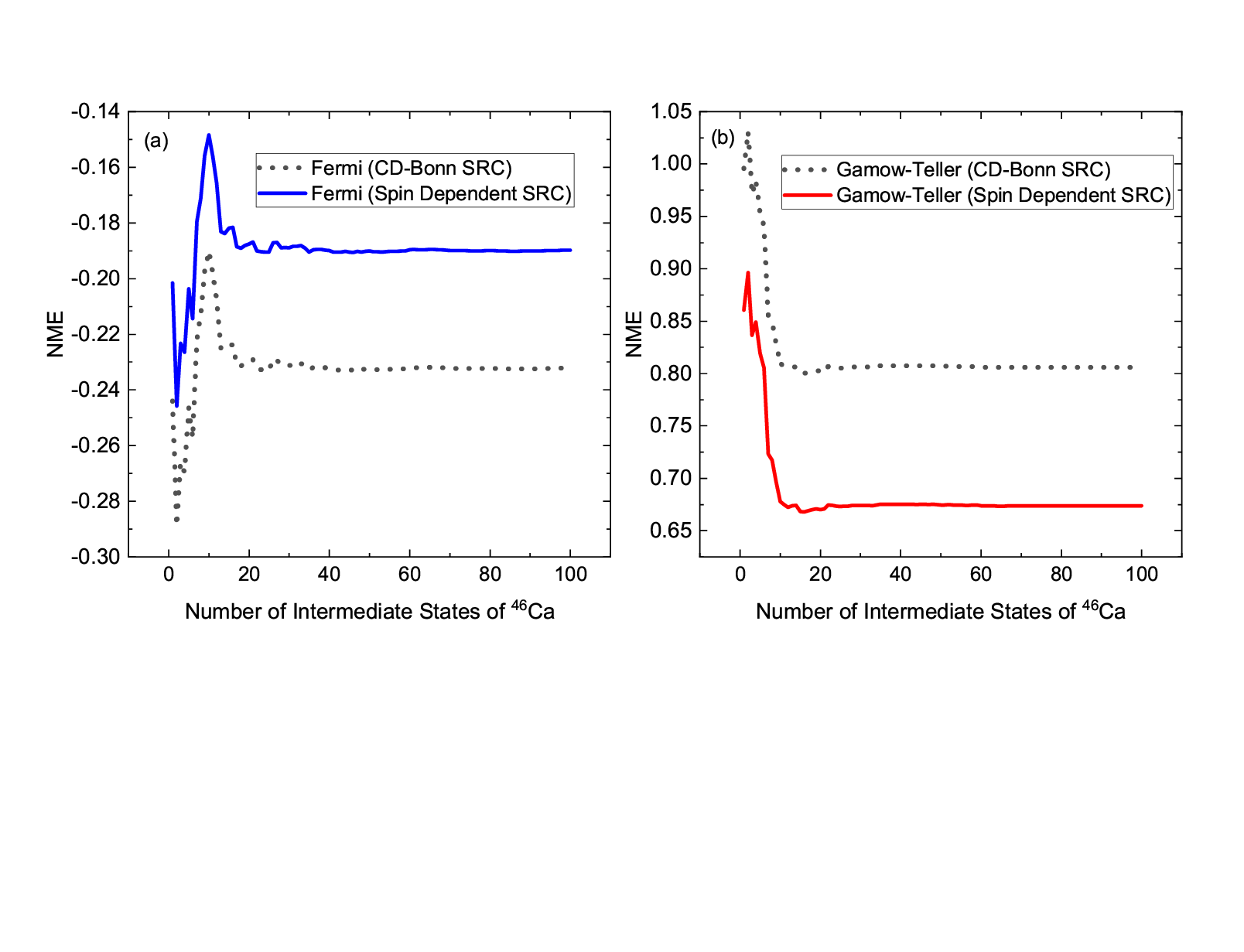}
\caption{\label{fig:nmevsnk}(Color online) The graph illustrates the variation of (a) Fermi and (b) Gamow-Teller NMEs for the $0\nu\beta\beta$ (the light neutrino-exchange mechanism) of $^{48}$Ca with the cutoff number of states ($N_c$) of the intermediate nucleus $^{46}$Ca (for TNA calculation). The NMEs are calculated using the GXPF1A Hamiltonian for CD-Bonn and spin-dependent SRC.}
\end{figure}
To assess the impact of the number of states on the calculated NMEs, we examine the dependence of the NMEs on the cutoff number of states ($N_c$) for each allowed $J_{m}^{\pi}$ of $^{46}$Ca, which acts as an intermediate nucleus for TNA calculation. The shell model code like KSHELL does not provide options on setting cut-off on excitation energy which may be more relevant, but it provides options to set cut-off on the number of states for each spin-parity of the intermediate states of $^{46}$Ca. Hence, we consider a certain number of intermediate states for each spin-parity of $^{46}$Ca with a cutoff according to the computational power available for the calculations. 
We express the NMEs as a function of $N_c$ in the closure method as
\begin{eqnarray}
{M}_{\alpha}=\sum_{J,N_{m}\leqslant N_c}{M}_{\alpha}(J^{\pi},N_{m})
\end{eqnarray}
where ${M}_{\alpha}(J^{\pi},N_m)$ is same as defined in Eq. (\ref{eq:nmepartial}). 
\begin{figure}
\includegraphics[trim=0cm 1cm 0cm 2cm,width=\linewidth]{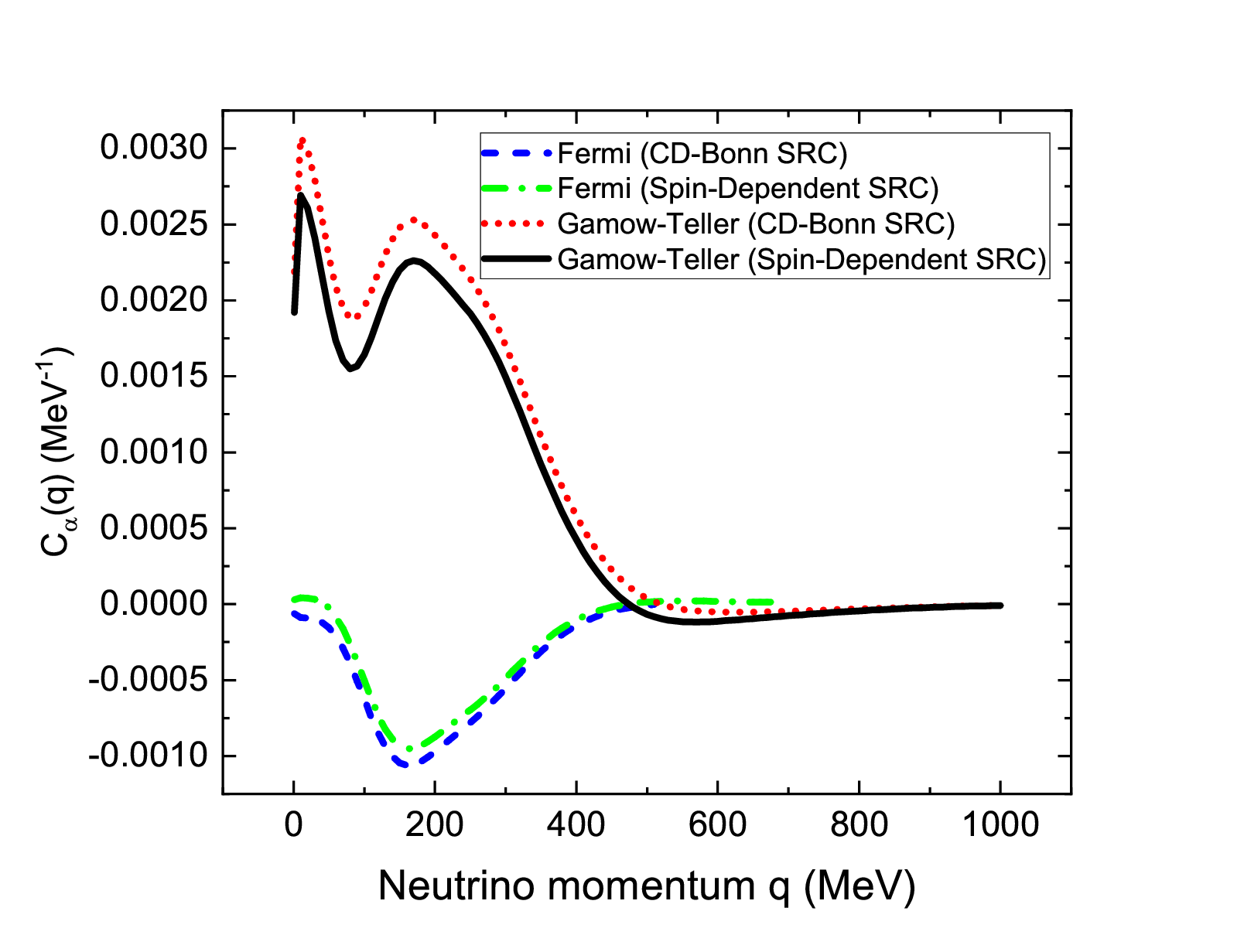}
\caption{\label{fig:NMEvsq}(Color online) Distribution of NMEs ($C_{\alpha}(q)$) with neutrino momentum ($q$) transfer. The NMEs are computed employing the closure method, considering both CD-Bonn SRC parametrization and spin-dependent SRC. These calculations are executed utilizing the GXPF1A Hamiltonian, with closure energy $\langle E\rangle$=0.5 MeV.}
\end{figure}
\begin{figure}
\includegraphics[trim=0cm 1cm 0cm 2cm,width=\linewidth]{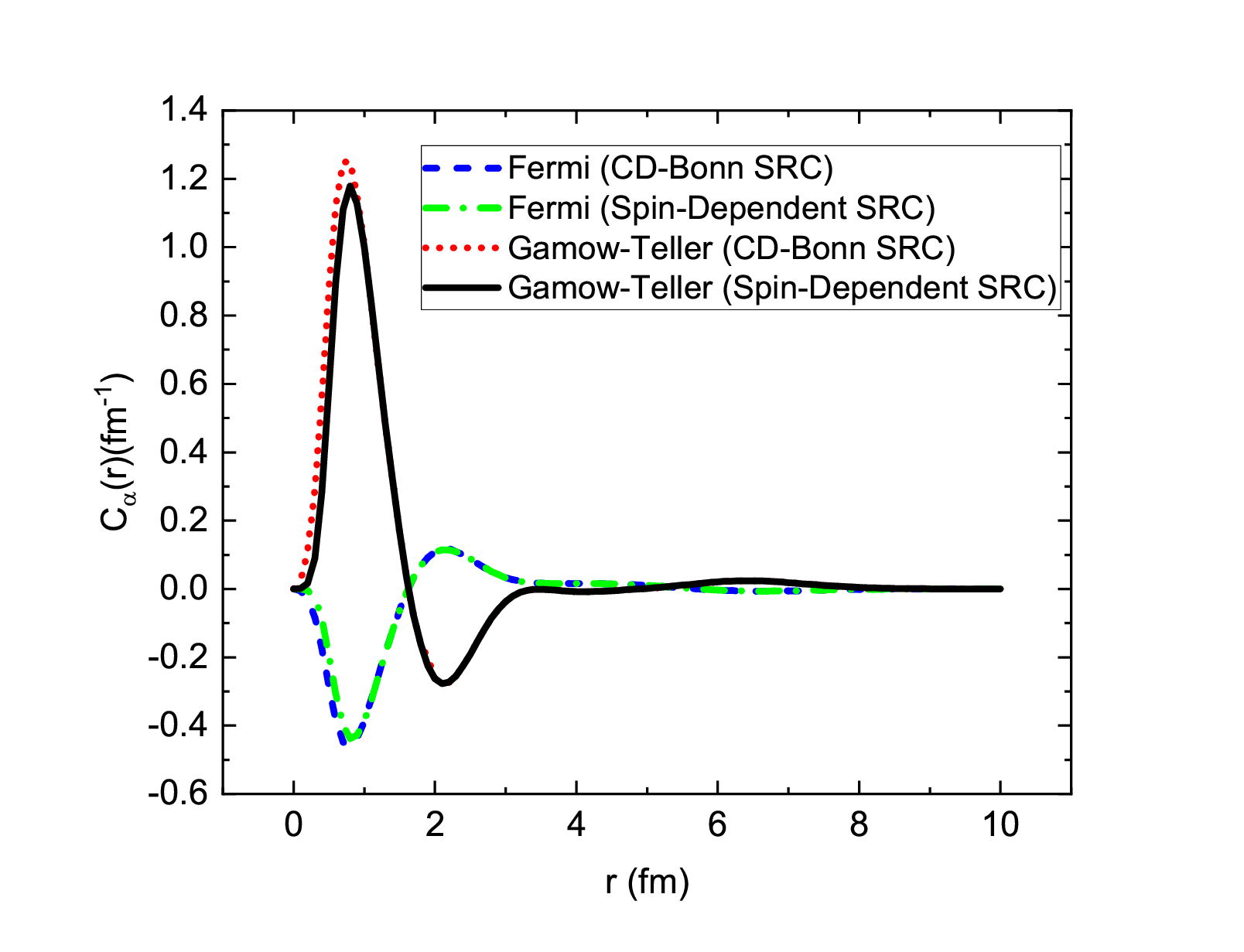}
\caption{\label{fig:NMEvsr}(Color online) Radial distribution of NME ($C_{\alpha}(r)$) with inter-nucleon distance ($r$) . The NMEs have been computed utilizing the closure method, considering both CD-Bonn SRC parametrization and spin-dependent SRC. These calculations are performed using the GXPF1A Hamiltonian, with a closure energy $\langle E\rangle$=0.5 MeV.}
\end{figure}
The dependence of the different types of NMEs on $N_c$ is shown in Fig.~\ref{fig:nmevsnk}. We compare the results for CD-Bonn SRC parametrization with spin-dependent SRC. We find that the first few low-lying states contribute constructively and destructively, but after $N_c=30$, the different types of NMEs reach a stable value. For larger $N_c$, the NMEs become mostly constant. To obtain NMEs with negligible uncertainty, we were able to consider $N_c=100$ for each allowed $J_m^{\pi}$ of $^{46}$Ca. We note that a similar dependence of NMEs on $N_c$ is seen for other SRC parametrizations. The overall pattern of dependence of NMEs on $N_c$ for spin-dependent SRC and CD-Bonn SRC is similar, but spin-dependent SRC NMEs are much smaller in magnitude as compared to NMEs for CD-Bonn type SRC. 
\subsection{\textbf{Neutrino momentum ($q$) and radial ($r$) distribution of NMEs}}

Further, we have examined the neutrino momentum transfer ($q$) distribution of NMEs. One can define $q$ dependent distribution ($C_{\alpha}(q)$) such that NME defined in Eq. (\ref{Eq:NMEMAIN}) can be written as \cite{menendez2017neutrinoless} 
\begin{equation}
    M_{\alpha}=\int_{0}^{\infty}C_{\alpha}(q)dq.
\end{equation}
Distribution of different $C_{\alpha}(q)$ is shown in Fig. \ref{fig:NMEvsq}. Here, NMEs are compared for CD-Bonn type SRC parametrization and spin-dependent SRC. It is found that most of the contributions of NME come from $q$ below 500 MeV. The peak contributions come from $q$ around 10 MeV and 180 MeV with peak value around 0.003 MeV$^{-1}$ for GT type NME with CD-Bonn SRC and peak value around 0.0027 MeV$^{-1}$ for GT type NME with spin-dependent SRC. The peak value for Fermi type NME is around -0.001 MeV$^{-1}$ for both CD-Bonn and spin-dependent SRC. The overall pattern of dependence of NMEs for both CD-Bonn and spin-dependent SRC is similar. 

The radial distribution of NME is also explored. One can write radial dependent NME distribution ($C_\alpha(r)$) \cite{PhysRevC.77.045503,menendez2008disassembling,menendez2017neutrinoless} such that 
\begin{equation}
    M_{\alpha}=\int_{0}^{\infty}C_{\alpha}(r)dr.
\end{equation}
Distribution of different $C_{\alpha}(r)$ is shown in Fig. \ref{fig:NMEvsr}. Here, NMEs are calculated using the closure method for CD-Bonn and spin-dependent SRC with $\langle E\rangle$=0.5 MeV. It is found that most of the contribution comes from $r$ less than 4 fm. The NMEs peak around 1 fm for both CD-Bonn and spin-dependent SRC. The maximum value $C_{\alpha}(r)$ is near 1.2 fm$^{-1}$ For GT and around -0.5 fm$^{-1}$ for Fermi type NME both CD-Bonn and spin-dependent SRC. Here also, the overall pattern of variation for CD-Bonn and spin-dependent SRC is similar, but they differ slightly in magnitude. 
\begin{figure}
\centering
\includegraphics[trim=0cm 1cm 0cm 0cm,width=\linewidth]{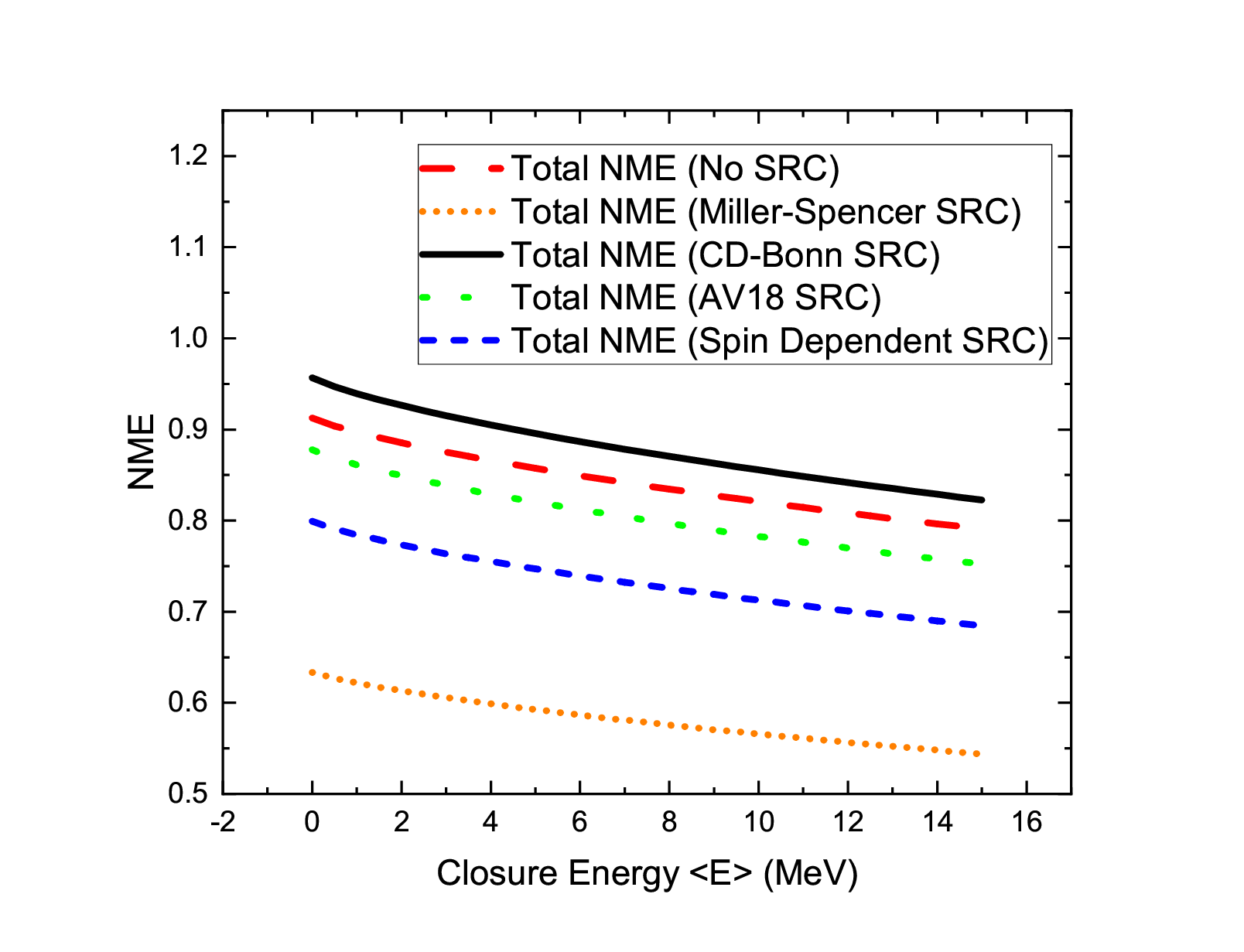}
\caption{\label{fig:nmevsclosure}(Color online) Variation in total NMEs pertaining to the $0\nu\beta\beta$ decay via the light neutrino-exchange mechanism in $^{48}$Ca, in relation to the closure energy $\langle E\rangle$. These computations employ the GXPF1A Hamiltonian and adopt different SRC parametrizations through the closure method. It's worth noting that the calculation of total NMEs does not encompass the tensor NMEs.}
\end{figure}
\subsection{\textbf{Variation of NME with closure energy}}

Finally, we have also shown the variation of different types of total NMEs with closure energy $\langle E\rangle$ in Fig. \ref{fig:nmevsclosure}. The NMEs shown here are the total NMEs in different parametrizations of the Jastrow-type approach and spin-dependent approach. In total NMEs, the Fermi and GT type NMEs are only considered, and tensor type NME is not included for the reason mentioned above.  For changing $\langle E\rangle$=0 to 15 MeV, there are about 10-15\% decrements of total NMEs in the closure method. Other than varying in magnitude, the spin-dependent SRC varies very similarly to other SRC parametrizations of the Jastrow-type approach.

\section{\label{sec:VI}Summary and conclusions}
In summary, our investigation has centered on comprehensively exploring the effects of spin-dependent SRC on the NMEs associated with the light neutrino exchange $0\nu\beta\beta$ decay of $^{48}$Ca. This exploration was juxtaposed against the backdrop of the traditional Jastrow-type approach to SRC, employing diverse parameterizations. All computations were performed within the framework of a nuclear shell model, incorporating the closure approximation and utilizing the GXPF1A Hamiltonian to encapsulate the complete $fp$ model space. The nuclear wave functions were derived through shell model diagonalization, leveraging the KSHELL code. These derived wave functions formed the bedrock for evaluating a multitude of TNA, integral to the formulation of the NMEs. The segment concerning TBME within the NMEs was computed through code developed by us.

Beyond the comparison of the NMEs of different types (Fermi, Gamow, and total) originating from spin-dependent SRC and Jastrow-type SRC, our analysis delved into the comparative dependencies of NMEs linked to attributes such as the coupled spin-parity of protons and neutrons, the number of intermediate states of TNA calculation, neutrino momentum, radial separation, and closure energy.

Our findings prominently highlight a discernible reduction in NMEs resulting from the application of spin-dependent SRCs, exhibiting, for instance, a 10-20\% decrease in NMEs when contrasted with Jastrow-type SRC with diverse parametrizations. This substantial variation may bear noticeable implications for the half-life and neutrino mass considerations. While exhibiting a different magnitude, the general pattern of variation in NMEs under the influence of spin-dependent SRC still bears a resemblance to that observed with the Jastrow-type approach.

In forthcoming research endeavors, our aim is to extend the influence of the spin-dependent SRC approach to encompass other candidates for $0\nu\beta\beta$ decay, exploring a spectrum of mechanisms and scenarios. 
\begin{acknowledgments}
S.S. is thankful to the Science and Engineering Research Board (SERB), Department of Science and Technology (DST), Government of India, for the gracious award of the SERB-National Postdoctoral Fellowship under Grant No. PDF/2022/003729. S.S. expresses profound gratitude to Prof. Rajdeep Chatterjee for the invaluable mentorship received throughout his tenure as a SERB-National Postdoctoral Fellow at the Indian Institute of Technology Roorkee.
\end{acknowledgments}
\nocite{}
\bibliography{main}

\begin{thebibliography}{59}%
\makeatletter
\providecommand \@ifxundefined [1]{%
 \@ifx{#1\undefined}
}%
\providecommand \@ifnum [1]{%
 \ifnum #1\expandafter \@firstoftwo
 \else \expandafter \@secondoftwo
 \fi
}%
\providecommand \@ifx [1]{%
 \ifx #1\expandafter \@firstoftwo
 \else \expandafter \@secondoftwo
 \fi
}%
\providecommand \natexlab [1]{#1}%
\providecommand \enquote  [1]{``#1''}%
\providecommand \bibnamefont  [1]{#1}%
\providecommand \bibfnamefont [1]{#1}%
\providecommand \citenamefont [1]{#1}%
\providecommand \href@noop [0]{\@secondoftwo}%
\providecommand \href [0]{\begingroup \@sanitize@url \@href}%
\providecommand \@href[1]{\@@startlink{#1}\@@href}%
\providecommand \@@href[1]{\endgroup#1\@@endlink}%
\providecommand \@sanitize@url [0]{\catcode `\\12\catcode `\$12\catcode
  `\&12\catcode `\#12\catcode `\^12\catcode `\_12\catcode `\%12\relax}%
\providecommand \@@startlink[1]{}%
\providecommand \@@endlink[0]{}%
\providecommand \url  [0]{\begingroup\@sanitize@url \@url }%
\providecommand \@url [1]{\endgroup\@href {#1}{\urlprefix }}%
\providecommand \urlprefix  [0]{URL }%
\providecommand \Eprint [0]{\href }%
\providecommand \doibase [0]{https://doi.org/}%
\providecommand \selectlanguage [0]{\@gobble}%
\providecommand \bibinfo  [0]{\@secondoftwo}%
\providecommand \bibfield  [0]{\@secondoftwo}%
\providecommand \translation [1]{[#1]}%
\providecommand \BibitemOpen [0]{}%
\providecommand \bibitemStop [0]{}%
\providecommand \bibitemNoStop [0]{.\EOS\space}%
\providecommand \EOS [0]{\spacefactor3000\relax}%
\providecommand \BibitemShut  [1]{\csname bibitem#1\endcsname}%
\let\auto@bib@innerbib\@empty
\bibitem [{\citenamefont {Dolinski}\ \emph {et~al.}(2019)\citenamefont
  {Dolinski}, \citenamefont {Poon},\ and\ \citenamefont
  {Rodejohann}}]{Dolinski:2019nrj}%
  \BibitemOpen
  \bibfield  {author} {\bibinfo {author} {\bibfnamefont {M.~J.}\ \bibnamefont
  {Dolinski}}, \bibinfo {author} {\bibfnamefont {A.~W.~P.}\ \bibnamefont
  {Poon}}, and\ \bibinfo {author} {\bibfnamefont {W.}~\bibnamefont
  {Rodejohann}},\ }\bibfield  {title} {\bibinfo {title} {{Neutrinoless
  Double-Beta Decay: Status and Prospects}},\ }\href
  {https://doi.org/10.1146/annurev-nucl-101918-023407} {\bibfield  {journal}
  {\bibinfo  {journal} {Ann. Rev. Nucl. Part. Sci.}\ }\textbf {\bibinfo
  {volume} {69}},\ \bibinfo {pages} {219} (\bibinfo {year} {2019})},\ \Eprint
  {https://arxiv.org/abs/1902.04097} {arXiv:1902.04097 [nucl-ex]} \BibitemShut
  {NoStop}%
\bibitem [{\citenamefont {Vergados}\ \emph {et~al.}(2016)\citenamefont
  {Vergados}, \citenamefont {Ejiri},\ and\ \citenamefont
  {\v{S}imkovic}}]{Vergados:2016hso}%
  \BibitemOpen
  \bibfield  {author} {\bibinfo {author} {\bibfnamefont {J.~D.}\ \bibnamefont
  {Vergados}}, \bibinfo {author} {\bibfnamefont {H.}~\bibnamefont {Ejiri}},
  and\ \bibinfo {author} {\bibfnamefont {F.}~\bibnamefont {\v{S}imkovic}},\
  }\bibfield  {title} {\bibinfo {title} {{Neutrinoless double beta decay and
  neutrino mass}},\ }\href {https://doi.org/10.1142/S0218301316300071}
  {\bibfield  {journal} {\bibinfo  {journal} {Int. J. Mod. Phys. E}\ }\textbf
  {\bibinfo {volume} {25}},\ \bibinfo {pages} {1630007} (\bibinfo {year}
  {2016})},\ \Eprint {https://arxiv.org/abs/1612.02924} {arXiv:1612.02924
  [hep-ph]} \BibitemShut {NoStop}%
\bibitem [{\citenamefont {Engel$~$}\ and\ \citenamefont
  {Men{\'e}ndez}(2017)}]{engel2017status}%
  \BibitemOpen
  \bibfield  {author} {\bibinfo {author} {\bibfnamefont {J.}~\bibnamefont
  {Engel$~$}}and\ \bibinfo {author} {\bibfnamefont {J.}~\bibnamefont
  {Men{\'e}ndez}},\ }\bibfield  {title} {\bibinfo {title} {Status and future of
  nuclear matrix elements for neutrinoless double-beta decay: a review},\
  }\href@noop {} {\bibfield  {journal} {\bibinfo  {journal} {Reports on
  Progress in Physics}\ }\textbf {\bibinfo {volume} {80}},\ \bibinfo {pages}
  {046301} (\bibinfo {year} {2017})}\BibitemShut {NoStop}%
\bibitem [{\citenamefont {Avignone}\ \emph {et~al.}(2008)\citenamefont
  {Avignone}, \citenamefont {Elliott},\ and\ \citenamefont
  {Engel}}]{RevModPhys.80.481}%
  \BibitemOpen
  \bibfield  {author} {\bibinfo {author} {\bibfnamefont {F.~T.}\ \bibnamefont
  {Avignone}}, \bibinfo {author} {\bibfnamefont {S.~R.}\ \bibnamefont
  {Elliott}}, and\ \bibinfo {author} {\bibfnamefont {J.}~\bibnamefont
  {Engel}},\ }\bibfield  {title} {\bibinfo {title} {Double beta decay, majorana
  neutrinos, and neutrino mass},\ }\href
  {https://doi.org/10.1103/RevModPhys.80.481} {\bibfield  {journal} {\bibinfo
  {journal} {Rev. Mod. Phys.}\ }\textbf {\bibinfo {volume} {80}},\ \bibinfo
  {pages} {481} (\bibinfo {year} {2008})}\BibitemShut {NoStop}%
\bibitem [{\citenamefont {Furry}(1939)}]{PhysRev.56.1184}%
  \BibitemOpen
  \bibfield  {author} {\bibinfo {author} {\bibfnamefont {W.~H.}\ \bibnamefont
  {Furry}},\ }\bibfield  {title} {\bibinfo {title} {On transition probabilities
  in double beta-disintegration},\ }\href
  {https://doi.org/10.1103/PhysRev.56.1184} {\bibfield  {journal} {\bibinfo
  {journal} {Phys. Rev.}\ }\textbf {\bibinfo {volume} {56}},\ \bibinfo {pages}
  {1184} (\bibinfo {year} {1939})}\BibitemShut {NoStop}%
\bibitem [{\citenamefont {Vergados}\ \emph {et~al.}(2012)\citenamefont
  {Vergados}, \citenamefont {Ejiri},\ and\ \citenamefont
  {{\v{S}}imkovic}}]{vergados2012theory}%
  \BibitemOpen
  \bibfield  {author} {\bibinfo {author} {\bibfnamefont {J.~D.}\ \bibnamefont
  {Vergados}}, \bibinfo {author} {\bibfnamefont {H.}~\bibnamefont {Ejiri}},
  and\ \bibinfo {author} {\bibfnamefont {F.}~\bibnamefont {{\v{S}}imkovic}},\
  }\bibfield  {title} {\bibinfo {title} {Theory of neutrinoless double-beta
  decay},\ }\href@noop {} {\bibfield  {journal} {\bibinfo  {journal} {Reports
  on Progress in Physics}\ }\textbf {\bibinfo {volume} {75}},\ \bibinfo {pages}
  {106301} (\bibinfo {year} {2012})}\BibitemShut {NoStop}%
\bibitem [{\citenamefont {Majorana}(1937)}]{majorana1937teoria}%
  \BibitemOpen
  \bibfield  {author} {\bibinfo {author} {\bibfnamefont {E.}~\bibnamefont
  {Majorana}},\ }\bibfield  {title} {\bibinfo {title} {Symmetric theory of
  electron and positron},\ }\href@noop {} {\bibfield  {journal} {\bibinfo
  {journal} {Il Nuovo Cimento (1924-1942)}\ }\textbf {\bibinfo {volume} {14}},\
  \bibinfo {pages} {171} (\bibinfo {year} {1937})}\BibitemShut {NoStop}%
\bibitem [{\citenamefont {Racah}(1937)}]{racah1937sulla}%
  \BibitemOpen
  \bibfield  {author} {\bibinfo {author} {\bibfnamefont {G.}~\bibnamefont
  {Racah}},\ }\bibfield  {title} {\bibinfo {title} {Sulla simmetria tra
  particelle e antiparticelle},\ }\href@noop {} {\bibfield  {journal} {\bibinfo
   {journal} {Il Nuovo Cimento}\ }\textbf {\bibinfo {volume} {14}},\ \bibinfo
  {pages} {322} (\bibinfo {year} {1937})}\BibitemShut {NoStop}%
\bibitem [{\citenamefont {Deppisch}\ \emph {et~al.}(2012)\citenamefont
  {Deppisch}, \citenamefont {Hirsch},\ and\ \citenamefont
  {P{\"a}s}}]{deppisch2012neutrinoless}%
  \BibitemOpen
  \bibfield  {author} {\bibinfo {author} {\bibfnamefont {F.~F.}\ \bibnamefont
  {Deppisch}}, \bibinfo {author} {\bibfnamefont {M.}~\bibnamefont {Hirsch}},
  and\ \bibinfo {author} {\bibfnamefont {H.}~\bibnamefont {P{\"a}s}},\
  }\bibfield  {title} {\bibinfo {title} {Neutrinoless double-beta decay and
  physics beyond the standard model},\ }\href@noop {} {\bibfield  {journal}
  {\bibinfo  {journal} {Journal of Physics G: Nuclear and Particle Physics}\
  }\textbf {\bibinfo {volume} {39}},\ \bibinfo {pages} {124007} (\bibinfo
  {year} {2012})}\BibitemShut {NoStop}%
\bibitem [{\citenamefont {Schechter$~$}\ and\ \citenamefont
  {Valle}(1982)}]{PhysRevD.25.2951}%
  \BibitemOpen
  \bibfield  {author} {\bibinfo {author} {\bibfnamefont {J.}~\bibnamefont
  {Schechter$~$}}and\ \bibinfo {author} {\bibfnamefont {J.~W.}\ \bibnamefont
  {Valle}},\ }\bibfield  {title} {\bibinfo {title} {Neutrinoless double-$\beta$
  decay in su (2)$\times$ u (1) theories},\ }\href@noop {} {\bibfield
  {journal} {\bibinfo  {journal} {Physical Review D}\ }\textbf {\bibinfo
  {volume} {25}},\ \bibinfo {pages} {2951} (\bibinfo {year}
  {1982})}\BibitemShut {NoStop}%
\bibitem [{\citenamefont {Rodejohann}(2011)}]{rodejohann2011neutrino}%
  \BibitemOpen
  \bibfield  {author} {\bibinfo {author} {\bibfnamefont {W.}~\bibnamefont
  {Rodejohann}},\ }\bibfield  {title} {\bibinfo {title} {Neutrino-less double
  beta decay and particle physics},\ }\href@noop {} {\bibfield  {journal}
  {\bibinfo  {journal} {International Journal of Modern Physics E}\ }\textbf
  {\bibinfo {volume} {20}},\ \bibinfo {pages} {1833} (\bibinfo {year}
  {2011})}\BibitemShut {NoStop}%
\bibitem [{\citenamefont {Aker}\ \emph {et~al.}(2022)\citenamefont {Aker} \emph
  {et~al.}}]{KATRIN:2021uub}%
  \BibitemOpen
  \bibfield  {author} {\bibinfo {author} {\bibfnamefont {M.}~\bibnamefont
  {Aker}} \emph {et~al.} (\bibinfo {collaboration} {KATRIN}),\ }\bibfield
  {title} {\bibinfo {title} {{Direct neutrino-mass measurement with
  sub-electronvolt sensitivity}},\ }\href
  {https://doi.org/10.1038/s41567-021-01463-1} {\bibfield  {journal} {\bibinfo
  {journal} {Nature Phys.}\ }\textbf {\bibinfo {volume} {18}},\ \bibinfo
  {pages} {160} (\bibinfo {year} {2022})},\ \Eprint
  {https://arxiv.org/abs/2105.08533} {arXiv:2105.08533 [hep-ex]} \BibitemShut
  {NoStop}%
\bibitem [{\citenamefont {Abe}\ \emph {et~al.}(2023)\citenamefont {Abe} \emph
  {et~al.}}]{KamLAND-Zen:2022tow}%
  \BibitemOpen
  \bibfield  {author} {\bibinfo {author} {\bibfnamefont {S.}~\bibnamefont
  {Abe}} \emph {et~al.} (\bibinfo {collaboration} {KamLAND-Zen}),\ }\bibfield
  {title} {\bibinfo {title} {{Search for the Majorana Nature of Neutrinos in
  the Inverted Mass Ordering Region with KamLAND-Zen}},\ }\href
  {https://doi.org/10.1103/PhysRevLett.130.051801} {\bibfield  {journal}
  {\bibinfo  {journal} {Phys. Rev. Lett.}\ }\textbf {\bibinfo {volume} {130}},\
  \bibinfo {pages} {051801} (\bibinfo {year} {2023})},\ \Eprint
  {https://arxiv.org/abs/2203.02139} {arXiv:2203.02139 [hep-ex]} \BibitemShut
  {NoStop}%
\bibitem [{\citenamefont {Tomoda}(1991)}]{tomoda1991double}%
  \BibitemOpen
  \bibfield  {author} {\bibinfo {author} {\bibfnamefont {T.}~\bibnamefont
  {Tomoda}},\ }\bibfield  {title} {\bibinfo {title} {Double beta decay},\
  }\href@noop {} {\bibfield  {journal} {\bibinfo  {journal} {Reports on
  Progress in Physics}\ }\textbf {\bibinfo {volume} {54}},\ \bibinfo {pages}
  {53} (\bibinfo {year} {1991})}\BibitemShut {NoStop}%
\bibitem [{\citenamefont {Rodin}\ \emph {et~al.}(2006)\citenamefont {Rodin},
  \citenamefont {Faessler}, \citenamefont {{\v{S}}imkovic},\ and\ \citenamefont
  {Vogel}}]{rodin2006assessment}%
  \BibitemOpen
  \bibfield  {author} {\bibinfo {author} {\bibfnamefont {V.}~\bibnamefont
  {Rodin}}, \bibinfo {author} {\bibfnamefont {A.}~\bibnamefont {Faessler}},
  \bibinfo {author} {\bibfnamefont {F.}~\bibnamefont {{\v{S}}imkovic}}, and\
  \bibinfo {author} {\bibfnamefont {P.}~\bibnamefont {Vogel}},\ }\bibfield
  {title} {\bibinfo {title} {Assessment of uncertainties in qrpa
  0$\nu$$\beta$$\beta$-decay nuclear matrix elements},\ }\href@noop {}
  {\bibfield  {journal} {\bibinfo  {journal} {Nuclear Physics A}\ }\textbf
  {\bibinfo {volume} {766}},\ \bibinfo {pages} {107} (\bibinfo {year}
  {2006})}\BibitemShut {NoStop}%
\bibitem [{\citenamefont {\ifmmode~\check{S}\else \v{S}\fi{}imkovic}\ \emph
  {et~al.}(1999)\citenamefont {\ifmmode~\check{S}\else \v{S}\fi{}imkovic},
  \citenamefont {Pantis}, \citenamefont {Vergados},\ and\ \citenamefont
  {Faessler}}]{PhysRevC.60.055502}%
  \BibitemOpen
  \bibfield  {author} {\bibinfo {author} {\bibfnamefont {F.}~\bibnamefont
  {\ifmmode~\check{S}\else \v{S}\fi{}imkovic}}, \bibinfo {author}
  {\bibfnamefont {G.}~\bibnamefont {Pantis}}, \bibinfo {author} {\bibfnamefont
  {J.~D.}\ \bibnamefont {Vergados}}, and\ \bibinfo {author} {\bibfnamefont
  {A.}~\bibnamefont {Faessler}},\ }\bibfield  {title} {\bibinfo {title}
  {Additional nucleon current contributions to neutrinoless double
  $\ensuremath{\beta}$ decay},\ }\href
  {https://doi.org/10.1103/PhysRevC.60.055502} {\bibfield  {journal} {\bibinfo
  {journal} {Phys. Rev. C}\ }\textbf {\bibinfo {volume} {60}},\ \bibinfo
  {pages} {055502} (\bibinfo {year} {1999})}\BibitemShut {NoStop}%
\bibitem [{\citenamefont {Mohapatra$~$}\ and\ \citenamefont
  {Senjanovi\ifmmode~\acute{c}\else \'{c}\fi{}}(1980)}]{PhysRevLett.44.912}%
  \BibitemOpen
  \bibfield  {author} {\bibinfo {author} {\bibfnamefont {R.~N.}\ \bibnamefont
  {Mohapatra$~$}}and\ \bibinfo {author} {\bibfnamefont {G.}~\bibnamefont
  {Senjanovi\ifmmode~\acute{c}\else \'{c}\fi{}}},\ }\bibfield  {title}
  {\bibinfo {title} {Neutrino mass and spontaneous parity nonconservation},\
  }\href {https://doi.org/10.1103/PhysRevLett.44.912} {\bibfield  {journal}
  {\bibinfo  {journal} {Phys. Rev. Lett.}\ }\textbf {\bibinfo {volume} {44}},\
  \bibinfo {pages} {912} (\bibinfo {year} {1980})}\BibitemShut {NoStop}%
\bibitem [{\citenamefont {Mohapatra$~$}\ and\ \citenamefont
  {Vergados}(1981)}]{PhysRevLett.47.1713}%
  \BibitemOpen
  \bibfield  {author} {\bibinfo {author} {\bibfnamefont {R.~N.}\ \bibnamefont
  {Mohapatra$~$}}and\ \bibinfo {author} {\bibfnamefont {J.~D.}\ \bibnamefont
  {Vergados}},\ }\bibfield  {title} {\bibinfo {title} {New contribution to
  neutrinoless double beta decay in gauge models},\ }\href
  {https://doi.org/10.1103/PhysRevLett.47.1713} {\bibfield  {journal} {\bibinfo
   {journal} {Phys. Rev. Lett.}\ }\textbf {\bibinfo {volume} {47}},\ \bibinfo
  {pages} {1713} (\bibinfo {year} {1981})}\BibitemShut {NoStop}%
\bibitem [{\citenamefont {Mohapatra}(1986)}]{PhysRevD.34.3457}%
  \BibitemOpen
  \bibfield  {author} {\bibinfo {author} {\bibfnamefont {R.~N.}\ \bibnamefont
  {Mohapatra}},\ }\bibfield  {title} {\bibinfo {title} {New contributions to
  neutrinoless double-beta decay in supersymmetric theories},\ }\href
  {https://doi.org/10.1103/PhysRevD.34.3457} {\bibfield  {journal} {\bibinfo
  {journal} {Phys. Rev. D}\ }\textbf {\bibinfo {volume} {34}},\ \bibinfo
  {pages} {3457} (\bibinfo {year} {1986})}\BibitemShut {NoStop}%
\bibitem [{\citenamefont {Vergados}(1987)}]{vergados1987neutrinoless}%
  \BibitemOpen
  \bibfield  {author} {\bibinfo {author} {\bibfnamefont {J.}~\bibnamefont
  {Vergados}},\ }\bibfield  {title} {\bibinfo {title} {Neutrinoless double
  $\beta$-decay without majorana neutrinos in supersymmetric theories},\
  }\href@noop {} {\bibfield  {journal} {\bibinfo  {journal} {Physics Letters
  B}\ }\textbf {\bibinfo {volume} {184}},\ \bibinfo {pages} {55} (\bibinfo
  {year} {1987})}\BibitemShut {NoStop}%
\bibitem [{\citenamefont {Caurier}\ \emph {et~al.}(2008)\citenamefont
  {Caurier}, \citenamefont {Men\'endez}, \citenamefont {Nowacki},\ and\
  \citenamefont {Poves}}]{PhysRevLett.100.052503}%
  \BibitemOpen
  \bibfield  {author} {\bibinfo {author} {\bibfnamefont {E.}~\bibnamefont
  {Caurier}}, \bibinfo {author} {\bibfnamefont {J.}~\bibnamefont {Men\'endez}},
  \bibinfo {author} {\bibfnamefont {F.}~\bibnamefont {Nowacki}}, and\ \bibinfo
  {author} {\bibfnamefont {A.}~\bibnamefont {Poves}},\ }\bibfield  {title}
  {\bibinfo {title} {Influence of pairing on the nuclear matrix elements of the
  neutrinoless $\ensuremath{\beta}\ensuremath{\beta}$ decays},\ }\href
  {https://doi.org/10.1103/PhysRevLett.100.052503} {\bibfield  {journal}
  {\bibinfo  {journal} {Phys. Rev. Lett.}\ }\textbf {\bibinfo {volume} {100}},\
  \bibinfo {pages} {052503} (\bibinfo {year} {2008})}\BibitemShut {NoStop}%
\bibitem [{\citenamefont {Horoi$~$}\ and\ \citenamefont
  {Stoica}(2010)}]{PhysRevC.81.024321}%
  \BibitemOpen
  \bibfield  {author} {\bibinfo {author} {\bibfnamefont {M.}~\bibnamefont
  {Horoi$~$}}and\ \bibinfo {author} {\bibfnamefont {S.}~\bibnamefont
  {Stoica}},\ }\bibfield  {title} {\bibinfo {title} {Shell model analysis of
  the neutrinoless double-$\ensuremath{\beta}$ decay of $^{48}\mathrm{Ca}$},\
  }\href {https://doi.org/10.1103/PhysRevC.81.024321} {\bibfield  {journal}
  {\bibinfo  {journal} {Phys. Rev. C}\ }\textbf {\bibinfo {volume} {81}},\
  \bibinfo {pages} {024321} (\bibinfo {year} {2010})}\BibitemShut {NoStop}%
\bibitem [{\citenamefont {Sen'kov$~$}\ and\ \citenamefont
  {Horoi}(2013)}]{PhysRevC.88.064312}%
  \BibitemOpen
  \bibfield  {author} {\bibinfo {author} {\bibfnamefont {R.~A.}\ \bibnamefont
  {Sen'kov$~$}}and\ \bibinfo {author} {\bibfnamefont {M.}~\bibnamefont
  {Horoi}},\ }\bibfield  {title} {\bibinfo {title} {Neutrinoless
  double-$\ensuremath{\beta}$ decay of ${}^{48}$ca in the shell model: Closure
  versus nonclosure approximation},\ }\href
  {https://doi.org/10.1103/PhysRevC.88.064312} {\bibfield  {journal} {\bibinfo
  {journal} {Phys. Rev. C}\ }\textbf {\bibinfo {volume} {88}},\ \bibinfo
  {pages} {064312} (\bibinfo {year} {2013})}\BibitemShut {NoStop}%
\bibitem [{\citenamefont {Brown}\ \emph {et~al.}(2014)\citenamefont {Brown},
  \citenamefont {Horoi},\ and\ \citenamefont
  {Sen'kov}}]{PhysRevLett.113.262501}%
  \BibitemOpen
  \bibfield  {author} {\bibinfo {author} {\bibfnamefont {B.~A.}\ \bibnamefont
  {Brown}}, \bibinfo {author} {\bibfnamefont {M.}~\bibnamefont {Horoi}}, and\
  \bibinfo {author} {\bibfnamefont {R.~A.}\ \bibnamefont {Sen'kov}},\
  }\bibfield  {title} {\bibinfo {title} {Nuclear structure aspects of
  neutrinoless double-$\ensuremath{\beta}$ decay},\ }\href
  {https://doi.org/10.1103/PhysRevLett.113.262501} {\bibfield  {journal}
  {\bibinfo  {journal} {Phys. Rev. Lett.}\ }\textbf {\bibinfo {volume} {113}},\
  \bibinfo {pages} {262501} (\bibinfo {year} {2014})}\BibitemShut {NoStop}%
\bibitem [{\citenamefont {Iwata}\ \emph {et~al.}(2016)\citenamefont {Iwata},
  \citenamefont {Shimizu}, \citenamefont {Otsuka}, \citenamefont {Utsuno},
  \citenamefont {Men\'endez}, \citenamefont {Honma},\ and\ \citenamefont
  {Abe}}]{PhysRevLett.116.112502}%
  \BibitemOpen
  \bibfield  {author} {\bibinfo {author} {\bibfnamefont {Y.}~\bibnamefont
  {Iwata}}, \bibinfo {author} {\bibfnamefont {N.}~\bibnamefont {Shimizu}},
  \bibinfo {author} {\bibfnamefont {T.}~\bibnamefont {Otsuka}}, \bibinfo
  {author} {\bibfnamefont {Y.}~\bibnamefont {Utsuno}}, \bibinfo {author}
  {\bibfnamefont {J.}~\bibnamefont {Men\'endez}}, \bibinfo {author}
  {\bibfnamefont {M.}~\bibnamefont {Honma}}, and\ \bibinfo {author}
  {\bibfnamefont {T.}~\bibnamefont {Abe}},\ }\bibfield  {title} {\bibinfo
  {title} {Large-scale shell-model analysis of the neutrinoless
  $\ensuremath{\beta}\ensuremath{\beta}$ decay of $^{48}\mathrm{Ca}$},\ }\href
  {https://doi.org/10.1103/PhysRevLett.116.112502} {\bibfield  {journal}
  {\bibinfo  {journal} {Phys. Rev. Lett.}\ }\textbf {\bibinfo {volume} {116}},\
  \bibinfo {pages} {112502} (\bibinfo {year} {2016})}\BibitemShut {NoStop}%
\bibitem [{\citenamefont {Barea$~$}\ and\ \citenamefont
  {Iachello}(2009)}]{PhysRevC.79.044301}%
  \BibitemOpen
  \bibfield  {author} {\bibinfo {author} {\bibfnamefont {J.}~\bibnamefont
  {Barea$~$}}and\ \bibinfo {author} {\bibfnamefont {F.}~\bibnamefont
  {Iachello}},\ }\bibfield  {title} {\bibinfo {title} {Neutrinoless
  double-$\ensuremath{\beta}$ decay in the microscopic interacting boson
  model},\ }\href {https://doi.org/10.1103/PhysRevC.79.044301} {\bibfield
  {journal} {\bibinfo  {journal} {Phys. Rev. C}\ }\textbf {\bibinfo {volume}
  {79}},\ \bibinfo {pages} {044301} (\bibinfo {year} {2009})}\BibitemShut
  {NoStop}%
\bibitem [{\citenamefont {Barea}\ \emph {et~al.}(2012)\citenamefont {Barea},
  \citenamefont {Kotila},\ and\ \citenamefont
  {Iachello}}]{PhysRevLett.109.042501}%
  \BibitemOpen
  \bibfield  {author} {\bibinfo {author} {\bibfnamefont {J.}~\bibnamefont
  {Barea}}, \bibinfo {author} {\bibfnamefont {J.}~\bibnamefont {Kotila}}, and\
  \bibinfo {author} {\bibfnamefont {F.}~\bibnamefont {Iachello}},\ }\bibfield
  {title} {\bibinfo {title} {Limits on neutrino masses from neutrinoless
  double-$\ensuremath{\beta}$ decay},\ }\href
  {https://doi.org/10.1103/PhysRevLett.109.042501} {\bibfield  {journal}
  {\bibinfo  {journal} {Phys. Rev. Lett.}\ }\textbf {\bibinfo {volume} {109}},\
  \bibinfo {pages} {042501} (\bibinfo {year} {2012})}\BibitemShut {NoStop}%
\bibitem [{\citenamefont {Rodr\'{\i}guez$~$}\ and\ \citenamefont
  {Mart\'{\i}nez-Pinedo}(2010)}]{PhysRevLett.105.252503}%
  \BibitemOpen
  \bibfield  {author} {\bibinfo {author} {\bibfnamefont {T.~R.}\ \bibnamefont
  {Rodr\'{\i}guez$~$}}and\ \bibinfo {author} {\bibfnamefont {G.}~\bibnamefont
  {Mart\'{\i}nez-Pinedo}},\ }\bibfield  {title} {\bibinfo {title} {Energy
  density functional study of nuclear matrix elements for neutrinoless
  $\ensuremath{\beta}\ensuremath{\beta}$ decay},\ }\href
  {https://doi.org/10.1103/PhysRevLett.105.252503} {\bibfield  {journal}
  {\bibinfo  {journal} {Phys. Rev. Lett.}\ }\textbf {\bibinfo {volume} {105}},\
  \bibinfo {pages} {252503} (\bibinfo {year} {2010})}\BibitemShut {NoStop}%
\bibitem [{\citenamefont {Song}\ \emph {et~al.}(2014)\citenamefont {Song},
  \citenamefont {Yao}, \citenamefont {Ring},\ and\ \citenamefont
  {Meng}}]{PhysRevC.90.054309}%
  \BibitemOpen
  \bibfield  {author} {\bibinfo {author} {\bibfnamefont {L.~S.}\ \bibnamefont
  {Song}}, \bibinfo {author} {\bibfnamefont {J.~M.}\ \bibnamefont {Yao}},
  \bibinfo {author} {\bibfnamefont {P.}~\bibnamefont {Ring}}, and\ \bibinfo
  {author} {\bibfnamefont {J.}~\bibnamefont {Meng}},\ }\bibfield  {title}
  {\bibinfo {title} {Relativistic description of nuclear matrix elements in
  neutrinoless double-$\ensuremath{\beta}$ decay},\ }\href
  {https://doi.org/10.1103/PhysRevC.90.054309} {\bibfield  {journal} {\bibinfo
  {journal} {Phys. Rev. C}\ }\textbf {\bibinfo {volume} {90}},\ \bibinfo
  {pages} {054309} (\bibinfo {year} {2014})}\BibitemShut {NoStop}%
\bibitem [{\citenamefont {Yao}\ \emph {et~al.}(2015)\citenamefont {Yao},
  \citenamefont {Song}, \citenamefont {Hagino}, \citenamefont {Ring},\ and\
  \citenamefont {Meng}}]{PhysRevC.91.024316}%
  \BibitemOpen
  \bibfield  {author} {\bibinfo {author} {\bibfnamefont {J.~M.}\ \bibnamefont
  {Yao}}, \bibinfo {author} {\bibfnamefont {L.~S.}\ \bibnamefont {Song}},
  \bibinfo {author} {\bibfnamefont {K.}~\bibnamefont {Hagino}}, \bibinfo
  {author} {\bibfnamefont {P.}~\bibnamefont {Ring}}, and\ \bibinfo {author}
  {\bibfnamefont {J.}~\bibnamefont {Meng}},\ }\bibfield  {title} {\bibinfo
  {title} {Systematic study of nuclear matrix elements in neutrinoless
  double-$\ensuremath{\beta}$ decay with a beyond-mean-field covariant density
  functional theory},\ }\href {https://doi.org/10.1103/PhysRevC.91.024316}
  {\bibfield  {journal} {\bibinfo  {journal} {Phys. Rev. C}\ }\textbf {\bibinfo
  {volume} {91}},\ \bibinfo {pages} {024316} (\bibinfo {year}
  {2015})}\BibitemShut {NoStop}%
\bibitem [{\citenamefont {Rath}\ \emph {et~al.}(2010)\citenamefont {Rath},
  \citenamefont {Chandra}, \citenamefont {Chaturvedi}, \citenamefont {Raina},\
  and\ \citenamefont {Hirsch}}]{PhysRevC.82.064310}%
  \BibitemOpen
  \bibfield  {author} {\bibinfo {author} {\bibfnamefont {P.~K.}\ \bibnamefont
  {Rath}}, \bibinfo {author} {\bibfnamefont {R.}~\bibnamefont {Chandra}},
  \bibinfo {author} {\bibfnamefont {K.}~\bibnamefont {Chaturvedi}}, \bibinfo
  {author} {\bibfnamefont {P.~K.}\ \bibnamefont {Raina}}, and\ \bibinfo
  {author} {\bibfnamefont {J.~G.}\ \bibnamefont {Hirsch}},\ }\bibfield  {title}
  {\bibinfo {title} {Uncertainties in nuclear transition matrix elements for
  neutrinoless $\ensuremath{\beta}\ensuremath{\beta}$ decay within the
  projected-hartree-fock-bogoliubov model},\ }\href
  {https://doi.org/10.1103/PhysRevC.82.064310} {\bibfield  {journal} {\bibinfo
  {journal} {Phys. Rev. C}\ }\textbf {\bibinfo {volume} {82}},\ \bibinfo
  {pages} {064310} (\bibinfo {year} {2010})}\BibitemShut {NoStop}%
\bibitem [{\citenamefont {Pastore}\ \emph {et~al.}(2018)\citenamefont
  {Pastore}, \citenamefont {Carlson}, \citenamefont {Cirigliano}, \citenamefont
  {Dekens}, \citenamefont {Mereghetti},\ and\ \citenamefont
  {Wiringa}}]{PhysRevC.97.014606}%
  \BibitemOpen
  \bibfield  {author} {\bibinfo {author} {\bibfnamefont {S.}~\bibnamefont
  {Pastore}}, \bibinfo {author} {\bibfnamefont {J.}~\bibnamefont {Carlson}},
  \bibinfo {author} {\bibfnamefont {V.}~\bibnamefont {Cirigliano}}, \bibinfo
  {author} {\bibfnamefont {W.}~\bibnamefont {Dekens}}, \bibinfo {author}
  {\bibfnamefont {E.}~\bibnamefont {Mereghetti}}, and\ \bibinfo {author}
  {\bibfnamefont {R.~B.}\ \bibnamefont {Wiringa}},\ }\bibfield  {title}
  {\bibinfo {title} {Neutrinoless double-$\ensuremath{\beta}$ decay matrix
  elements in light nuclei},\ }\href
  {https://doi.org/10.1103/PhysRevC.97.014606} {\bibfield  {journal} {\bibinfo
  {journal} {Phys. Rev. C}\ }\textbf {\bibinfo {volume} {97}},\ \bibinfo
  {pages} {014606} (\bibinfo {year} {2018})}\BibitemShut {NoStop}%
\bibitem [{\citenamefont {Wang}\ \emph {et~al.}(2019)\citenamefont {Wang},
  \citenamefont {Hayes}, \citenamefont {Carlson}, \citenamefont {Dong},
  \citenamefont {Mereghetti}, \citenamefont {Pastore},\ and\ \citenamefont
  {Wiringa}}]{wang2019comparison}%
  \BibitemOpen
  \bibfield  {author} {\bibinfo {author} {\bibfnamefont {X.~B.}\ \bibnamefont
  {Wang}}, \bibinfo {author} {\bibfnamefont {A.}~\bibnamefont {Hayes}},
  \bibinfo {author} {\bibfnamefont {J.}~\bibnamefont {Carlson}}, \bibinfo
  {author} {\bibfnamefont {G.}~\bibnamefont {Dong}}, \bibinfo {author}
  {\bibfnamefont {E.}~\bibnamefont {Mereghetti}}, \bibinfo {author}
  {\bibfnamefont {S.}~\bibnamefont {Pastore}}, and\ \bibinfo {author}
  {\bibfnamefont {R.~B.}\ \bibnamefont {Wiringa}},\ }\bibfield  {title}
  {\bibinfo {title} {Comparison between variational monte carlo and shell model
  calculations of neutrinoless double beta decay matrix elements in light
  nuclei},\ }\href@noop {} {\bibfield  {journal} {\bibinfo  {journal} {Physics
  Letters B}\ }\textbf {\bibinfo {volume} {798}},\ \bibinfo {pages} {134974}
  (\bibinfo {year} {2019})}\BibitemShut {NoStop}%
\bibitem [{\citenamefont {Cirigliano}\ \emph {et~al.}(2019)\citenamefont
  {Cirigliano}, \citenamefont {Dekens}, \citenamefont {de~Vries}, \citenamefont
  {Graesser}, \citenamefont {Mereghetti}, \citenamefont {Pastore},
  \citenamefont {Piarulli}, \citenamefont {van Kolck},\ and\ \citenamefont
  {Wiringa}}]{PhysRevC.100.055504}%
  \BibitemOpen
  \bibfield  {author} {\bibinfo {author} {\bibfnamefont {V.}~\bibnamefont
  {Cirigliano}}, \bibinfo {author} {\bibfnamefont {W.}~\bibnamefont {Dekens}},
  \bibinfo {author} {\bibfnamefont {J.}~\bibnamefont {de~Vries}}, \bibinfo
  {author} {\bibfnamefont {M.~L.}\ \bibnamefont {Graesser}}, \bibinfo {author}
  {\bibfnamefont {E.}~\bibnamefont {Mereghetti}}, \bibinfo {author}
  {\bibfnamefont {S.}~\bibnamefont {Pastore}}, \bibinfo {author} {\bibfnamefont
  {M.}~\bibnamefont {Piarulli}}, \bibinfo {author} {\bibfnamefont
  {U.}~\bibnamefont {van Kolck}}, and\ \bibinfo {author} {\bibfnamefont
  {R.~B.}\ \bibnamefont {Wiringa}},\ }\bibfield  {title} {\bibinfo {title}
  {Renormalized approach to neutrinoless double-$\ensuremath{\beta}$ decay},\
  }\href {https://doi.org/10.1103/PhysRevC.100.055504} {\bibfield  {journal}
  {\bibinfo  {journal} {Phys. Rev. C}\ }\textbf {\bibinfo {volume} {100}},\
  \bibinfo {pages} {055504} (\bibinfo {year} {2019})}\BibitemShut {NoStop}%
\bibitem [{\citenamefont {Sarkar}\ \emph
  {et~al.}(2020{\natexlab{a}})\citenamefont {Sarkar}, \citenamefont {Iwata},\
  and\ \citenamefont {Raina}}]{PhysRevC.102.034317}%
  \BibitemOpen
  \bibfield  {author} {\bibinfo {author} {\bibfnamefont {S.}~\bibnamefont
  {Sarkar}}, \bibinfo {author} {\bibfnamefont {Y.}~\bibnamefont {Iwata}}, and\
  \bibinfo {author} {\bibfnamefont {P.~K.}\ \bibnamefont {Raina}},\ }\bibfield
  {title} {\bibinfo {title} {Nuclear matrix elements for the
  $\ensuremath{\lambda}$ mechanism of
  $0\ensuremath{\nu}\ensuremath{\beta}\ensuremath{\beta}$ decay of
  $^{48}\mathrm{Ca}$ in the nuclear shell-model: Closure versus nonclosure
  approach},\ }\href {https://doi.org/10.1103/PhysRevC.102.034317} {\bibfield
  {journal} {\bibinfo  {journal} {Phys. Rev. C}\ }\textbf {\bibinfo {volume}
  {102}},\ \bibinfo {pages} {034317} (\bibinfo {year}
  {2020}{\natexlab{a}})}\BibitemShut {NoStop}%
\bibitem [{\citenamefont {Sarkar}\ \emph
  {et~al.}(2020{\natexlab{b}})\citenamefont {Sarkar}, \citenamefont {Kumar},
  \citenamefont {Jha},\ and\ \citenamefont {Raina}}]{PhysRevC.101.014307}%
  \BibitemOpen
  \bibfield  {author} {\bibinfo {author} {\bibfnamefont {S.}~\bibnamefont
  {Sarkar}}, \bibinfo {author} {\bibfnamefont {P.}~\bibnamefont {Kumar}},
  \bibinfo {author} {\bibfnamefont {K.}~\bibnamefont {Jha}}, and\ \bibinfo
  {author} {\bibfnamefont {P.~K.}\ \bibnamefont {Raina}},\ }\bibfield  {title}
  {\bibinfo {title} {Sensitivity of nuclear matrix elements of
  $0\ensuremath{\nu}\ensuremath{\beta}\ensuremath{\beta}$ of $^{48}\mathrm{Ca}$
  to different components of the two-nucleon interaction},\ }\href
  {https://doi.org/10.1103/PhysRevC.101.014307} {\bibfield  {journal} {\bibinfo
   {journal} {Phys. Rev. C}\ }\textbf {\bibinfo {volume} {101}},\ \bibinfo
  {pages} {014307} (\bibinfo {year} {2020}{\natexlab{b}})}\BibitemShut
  {NoStop}%
\bibitem [{\citenamefont {Ahmed}\ and\ \citenamefont
  {Horoi}(2020)}]{PhysRevC.101.035504}%
  \BibitemOpen
  \bibfield  {author} {\bibinfo {author} {\bibfnamefont {F.}~\bibnamefont
  {Ahmed}}and\ \bibinfo {author} {\bibfnamefont {M.}~\bibnamefont {Horoi}},\
  }\bibfield  {title} {\bibinfo {title} {Interference effects for
  $0\ensuremath{\nu}\ensuremath{\beta}\ensuremath{\beta}$ decay in the
  left-right symmetric model},\ }\href
  {https://doi.org/10.1103/PhysRevC.101.035504} {\bibfield  {journal} {\bibinfo
   {journal} {Phys. Rev. C}\ }\textbf {\bibinfo {volume} {101}},\ \bibinfo
  {pages} {035504} (\bibinfo {year} {2020})}\BibitemShut {NoStop}%
\bibitem [{\citenamefont {Horoi$~$}\ and\ \citenamefont
  {Neacsu}(2018)}]{PhysRevC.98.035502}%
  \BibitemOpen
  \bibfield  {author} {\bibinfo {author} {\bibfnamefont {M.}~\bibnamefont
  {Horoi$~$}}and\ \bibinfo {author} {\bibfnamefont {A.}~\bibnamefont
  {Neacsu}},\ }\bibfield  {title} {\bibinfo {title} {Shell model study of using
  an effective field theory for disentangling several contributions to
  neutrinoless double-$\ensuremath{\beta}$ decay},\ }\href
  {https://doi.org/10.1103/PhysRevC.98.035502} {\bibfield  {journal} {\bibinfo
  {journal} {Phys. Rev. C}\ }\textbf {\bibinfo {volume} {98}},\ \bibinfo
  {pages} {035502} (\bibinfo {year} {2018})}\BibitemShut {NoStop}%
\bibitem [{\citenamefont {Horoi}(2013)}]{PhysRevC.87.014320}%
  \BibitemOpen
  \bibfield  {author} {\bibinfo {author} {\bibfnamefont {M.}~\bibnamefont
  {Horoi}},\ }\bibfield  {title} {\bibinfo {title} {Shell model analysis of
  competing contributions to the double-$\ensuremath{\beta}$ decay of
  ${}^{48}$ca},\ }\href {https://doi.org/10.1103/PhysRevC.87.014320} {\bibfield
   {journal} {\bibinfo  {journal} {Phys. Rev. C}\ }\textbf {\bibinfo {volume}
  {87}},\ \bibinfo {pages} {014320} (\bibinfo {year} {2013})}\BibitemShut
  {NoStop}%
\bibitem [{\citenamefont {Neacsu}\ \emph
  {et~al.}(2012{\natexlab{a}})\citenamefont {Neacsu}, \citenamefont {Stoica},\
  and\ \citenamefont {Horoi}}]{PhysRevC.86.067304}%
  \BibitemOpen
  \bibfield  {author} {\bibinfo {author} {\bibfnamefont {A.}~\bibnamefont
  {Neacsu}}, \bibinfo {author} {\bibfnamefont {S.}~\bibnamefont {Stoica}}, and\
  \bibinfo {author} {\bibfnamefont {M.}~\bibnamefont {Horoi}},\ }\bibfield
  {title} {\bibinfo {title} {Fast, efficient calculations of the two-body
  matrix elements of the transition operators for neutrinoless double-$\beta$
  decay},\ }\href {https://doi.org/10.1103/PhysRevC.86.067304} {\bibfield
  {journal} {\bibinfo  {journal} {Phys. Rev. C}\ }\textbf {\bibinfo {volume}
  {86}},\ \bibinfo {pages} {067304} (\bibinfo {year}
  {2012}{\natexlab{a}})}\BibitemShut {NoStop}%
\bibitem [{\citenamefont {Menendez}\ \emph {et~al.}(2009)\citenamefont
  {Menendez}, \citenamefont {Poves}, \citenamefont {Caurier},\ and\
  \citenamefont {Nowacki}}]{Menendez:2008jp}%
  \BibitemOpen
  \bibfield  {author} {\bibinfo {author} {\bibfnamefont {J.}~\bibnamefont
  {Menendez}}, \bibinfo {author} {\bibfnamefont {A.}~\bibnamefont {Poves}},
  \bibinfo {author} {\bibfnamefont {E.}~\bibnamefont {Caurier}}, and\ \bibinfo
  {author} {\bibfnamefont {F.}~\bibnamefont {Nowacki}},\ }\bibfield  {title}
  {\bibinfo {title} {{Disassembling the Nuclear Matrix Elements of the
  Neutrinoless beta beta Decay}},\ }\href
  {https://doi.org/10.1016/j.nuclphysa.2008.12.005} {\bibfield  {journal}
  {\bibinfo  {journal} {Nucl. Phys. A}\ }\textbf {\bibinfo {volume} {818}},\
  \bibinfo {pages} {139} (\bibinfo {year} {2009})},\ \Eprint
  {https://arxiv.org/abs/0801.3760} {arXiv:0801.3760 [nucl-th]} \BibitemShut
  {NoStop}%
\bibitem [{\citenamefont {\ifmmode~\check{S}\else \v{S}\fi{}imkovic}\ \emph
  {et~al.}(2009)\citenamefont {\ifmmode~\check{S}\else \v{S}\fi{}imkovic},
  \citenamefont {Faessler}, \citenamefont {M\"uther}, \citenamefont {Rodin},\
  and\ \citenamefont {Stauf}}]{PhysRevC.79.055501}%
  \BibitemOpen
  \bibfield  {author} {\bibinfo {author} {\bibfnamefont {F.}~\bibnamefont
  {\ifmmode~\check{S}\else \v{S}\fi{}imkovic}}, \bibinfo {author}
  {\bibfnamefont {A.}~\bibnamefont {Faessler}}, \bibinfo {author}
  {\bibfnamefont {H.}~\bibnamefont {M\"uther}}, \bibinfo {author}
  {\bibfnamefont {V.}~\bibnamefont {Rodin}}, and\ \bibinfo {author}
  {\bibfnamefont {M.}~\bibnamefont {Stauf}},\ }\bibfield  {title} {\bibinfo
  {title} {$0\ensuremath{\nu}\ensuremath{\beta}\ensuremath{\beta}$-decay
  nuclear matrix elements with self-consistent short-range correlations},\
  }\href {https://doi.org/10.1103/PhysRevC.79.055501} {\bibfield  {journal}
  {\bibinfo  {journal} {Phys. Rev. C}\ }\textbf {\bibinfo {volume} {79}},\
  \bibinfo {pages} {055501} (\bibinfo {year} {2009})}\BibitemShut {NoStop}%
\bibitem [{\citenamefont {Vogel}(2012)}]{vogel2012nuclear}%
  \BibitemOpen
  \bibfield  {author} {\bibinfo {author} {\bibfnamefont {P.}~\bibnamefont
  {Vogel}},\ }\bibfield  {title} {\bibinfo {title} {Nuclear structure and
  double beta decay},\ }\href@noop {} {\bibfield  {journal} {\bibinfo
  {journal} {Journal of Physics G: Nuclear and Particle Physics}\ }\textbf
  {\bibinfo {volume} {39}},\ \bibinfo {pages} {124002} (\bibinfo {year}
  {2012})}\BibitemShut {NoStop}%
\bibitem [{\citenamefont {Feldmeier}\ \emph {et~al.}(1998)\citenamefont
  {Feldmeier}, \citenamefont {Neff}, \citenamefont {Roth},\ and\ \citenamefont
  {Schnack}}]{feldmeier1998unitary}%
  \BibitemOpen
  \bibfield  {author} {\bibinfo {author} {\bibfnamefont {H.}~\bibnamefont
  {Feldmeier}}, \bibinfo {author} {\bibfnamefont {T.}~\bibnamefont {Neff}},
  \bibinfo {author} {\bibfnamefont {R.}~\bibnamefont {Roth}}, and\ \bibinfo
  {author} {\bibfnamefont {J.}~\bibnamefont {Schnack}},\ }\bibfield  {title}
  {\bibinfo {title} {A unitary correlation operator method},\ }\href@noop {}
  {\bibfield  {journal} {\bibinfo  {journal} {Nuclear Physics A}\ }\textbf
  {\bibinfo {volume} {632}},\ \bibinfo {pages} {61} (\bibinfo {year}
  {1998})}\BibitemShut {NoStop}%
\bibitem [{\citenamefont {Neff$~$}\ and\ \citenamefont
  {Feldmeier}(2003)}]{NEFF2003311}%
  \BibitemOpen
  \bibfield  {author} {\bibinfo {author} {\bibfnamefont {T.}~\bibnamefont
  {Neff$~$}}and\ \bibinfo {author} {\bibfnamefont {H.}~\bibnamefont
  {Feldmeier}},\ }\bibfield  {title} {\bibinfo {title} {Tensor correlations in
  the unitary correlation operator method},\ }\href
  {https://doi.org/https://doi.org/10.1016/S0375-9474(02)01307-6} {\bibfield
  {journal} {\bibinfo  {journal} {Nuclear Physics A}\ }\textbf {\bibinfo
  {volume} {713}},\ \bibinfo {pages} {311 } (\bibinfo {year}
  {2003})}\BibitemShut {NoStop}%
\bibitem [{\citenamefont {Roth}\ \emph {et~al.}(2004)\citenamefont {Roth},
  \citenamefont {Neff}, \citenamefont {Hergert},\ and\ \citenamefont
  {Feldmeier}}]{ROTH20043}%
  \BibitemOpen
  \bibfield  {author} {\bibinfo {author} {\bibfnamefont {R.}~\bibnamefont
  {Roth}}, \bibinfo {author} {\bibfnamefont {T.}~\bibnamefont {Neff}}, \bibinfo
  {author} {\bibfnamefont {H.}~\bibnamefont {Hergert}}, and\ \bibinfo {author}
  {\bibfnamefont {H.}~\bibnamefont {Feldmeier}},\ }\bibfield  {title} {\bibinfo
  {title} {Nuclear structure based on correlated realistic nucleon–nucleon
  potentials},\ }\href
  {https://doi.org/https://doi.org/10.1016/j.nuclphysa.2004.08.024} {\bibfield
  {journal} {\bibinfo  {journal} {Nuclear Physics A}\ }\textbf {\bibinfo
  {volume} {745}},\ \bibinfo {pages} {3 } (\bibinfo {year} {2004})}\BibitemShut
  {NoStop}%
\bibitem [{\citenamefont {Benhar}\ \emph {et~al.}(2014)\citenamefont {Benhar},
  \citenamefont {Biondi},\ and\ \citenamefont {Speranza}}]{PhysRevC.90.065504}%
  \BibitemOpen
  \bibfield  {author} {\bibinfo {author} {\bibfnamefont {O.}~\bibnamefont
  {Benhar}}, \bibinfo {author} {\bibfnamefont {R.}~\bibnamefont {Biondi}}, and\
  \bibinfo {author} {\bibfnamefont {E.}~\bibnamefont {Speranza}},\ }\bibfield
  {title} {\bibinfo {title} {Short-range correlation effects on the nuclear
  matrix element of neutrinoless double-$\ensuremath{\beta}$ decay},\ }\href
  {https://doi.org/10.1103/PhysRevC.90.065504} {\bibfield  {journal} {\bibinfo
  {journal} {Phys. Rev. C}\ }\textbf {\bibinfo {volume} {90}},\ \bibinfo
  {pages} {065504} (\bibinfo {year} {2014})}\BibitemShut {NoStop}%
\bibitem [{\citenamefont {Kotila$~$}\ and\ \citenamefont
  {Iachello}(2012)}]{PhysRevC.85.034316}%
  \BibitemOpen
  \bibfield  {author} {\bibinfo {author} {\bibfnamefont {J.}~\bibnamefont
  {Kotila$~$}}and\ \bibinfo {author} {\bibfnamefont {F.}~\bibnamefont
  {Iachello}},\ }\bibfield  {title} {\bibinfo {title} {Phase-space factors for
  double-$\ensuremath{\beta}$ decay},\ }\href
  {https://doi.org/10.1103/PhysRevC.85.034316} {\bibfield  {journal} {\bibinfo
  {journal} {Phys. Rev. C}\ }\textbf {\bibinfo {volume} {85}},\ \bibinfo
  {pages} {034316} (\bibinfo {year} {2012})}\BibitemShut {NoStop}%
\bibitem [{\citenamefont {Neacsu}\ \emph
  {et~al.}(2012{\natexlab{b}})\citenamefont {Neacsu}, \citenamefont {Stoica},\
  and\ \citenamefont {Horoi}}]{neacsu2012fast}%
  \BibitemOpen
  \bibfield  {author} {\bibinfo {author} {\bibfnamefont {A.}~\bibnamefont
  {Neacsu}}, \bibinfo {author} {\bibfnamefont {S.}~\bibnamefont {Stoica}}, and\
  \bibinfo {author} {\bibfnamefont {M.}~\bibnamefont {Horoi}},\ }\bibfield
  {title} {\bibinfo {title} {Fast, efficient calculations of the two-body
  matrix elements of the transition operators for neutrinoless double-$\beta$
  decay},\ }\href@noop {} {\bibfield  {journal} {\bibinfo  {journal} {Physical
  Review C}\ }\textbf {\bibinfo {volume} {86}},\ \bibinfo {pages} {067304}
  (\bibinfo {year} {2012}{\natexlab{b}})}\BibitemShut {NoStop}%
\bibitem [{\citenamefont {e~Naturali}\ \emph {et~al.}()\citenamefont
  {e~Naturali}, \citenamefont {Speranza},\ and\ \citenamefont
  {Noccioli}}]{ecorrelation}%
  \BibitemOpen
  \bibfield  {author} {\bibinfo {author} {\bibfnamefont {F.}~\bibnamefont
  {e~Naturali}}, \bibinfo {author} {\bibfnamefont {E.}~\bibnamefont
  {Speranza}}, and\ \bibinfo {author} {\bibfnamefont {O.~B.}\ \bibnamefont
  {Noccioli}},\ }\bibfield  {title} {\bibinfo {title} {Correlation effects on
  the nuclear matrix element of neutrinoless double $\beta$-decay},\ }\href
  {http://chimera.roma1.infn.it/OMAR/personal/theses/speranza.pdf} {\
  }\BibitemShut {NoStop}%
\bibitem [{\citenamefont {Valli}(2007)}]{valli2007shear}%
  \BibitemOpen
  \bibfield  {author} {\bibinfo {author} {\bibfnamefont {M.}~\bibnamefont
  {Valli}},\ }\emph {\bibinfo {title} {Shear Viscosity of neutron matter from
  realistic nucleon-nucleon interactions}},\ \href@noop {} {Ph.D. thesis},\
  \bibinfo  {school} {Ph. D. thesis, Sapienza University of Rome} (\bibinfo
  {year} {2007})\BibitemShut {NoStop}%
\bibitem [{\citenamefont {Kortelainen}\ \emph {et~al.}(2007)\citenamefont
  {Kortelainen}, \citenamefont {Civitarese}, \citenamefont {Suhonen},\ and\
  \citenamefont {Toivanen}}]{kortelainen2007short}%
  \BibitemOpen
  \bibfield  {author} {\bibinfo {author} {\bibfnamefont {M.}~\bibnamefont
  {Kortelainen}}, \bibinfo {author} {\bibfnamefont {O.}~\bibnamefont
  {Civitarese}}, \bibinfo {author} {\bibfnamefont {J.}~\bibnamefont {Suhonen}},
  and\ \bibinfo {author} {\bibfnamefont {J.}~\bibnamefont {Toivanen}},\
  }\bibfield  {title} {\bibinfo {title} {Short-range correlations and
  neutrinoless double beta decay},\ }\href@noop {} {\bibfield  {journal}
  {\bibinfo  {journal} {Physics Letters B}\ }\textbf {\bibinfo {volume}
  {647}},\ \bibinfo {pages} {128} (\bibinfo {year} {2007})}\BibitemShut
  {NoStop}%
\bibitem [{\citenamefont {Kortelainen$~$}\ and\ \citenamefont
  {Suhonen}(2007)}]{PhysRevC.75.051303}%
  \BibitemOpen
  \bibfield  {author} {\bibinfo {author} {\bibfnamefont {M.}~\bibnamefont
  {Kortelainen$~$}}and\ \bibinfo {author} {\bibfnamefont {J.}~\bibnamefont
  {Suhonen}},\ }\bibfield  {title} {\bibinfo {title} {Improved short-range
  correlations and $0\ensuremath{\nu}\ensuremath{\beta}\ensuremath{\beta}$
  nuclear matrix elements of $^{76}\mathrm{Ge}$ and $^{82}\mathrm{Se}$},\
  }\href {https://doi.org/10.1103/PhysRevC.75.051303} {\bibfield  {journal}
  {\bibinfo  {journal} {Phys. Rev. C}\ }\textbf {\bibinfo {volume} {75}},\
  \bibinfo {pages} {051303} (\bibinfo {year} {2007})}\BibitemShut {NoStop}%
\bibitem [{\citenamefont {{\v{S}}imkovic}\ \emph {et~al.}(2009)\citenamefont
  {{\v{S}}imkovic}, \citenamefont {Faessler}, \citenamefont {M{\"u}ther},
  \citenamefont {Rodin},\ and\ \citenamefont {Stauf}}]{vsimkovic20090}%
  \BibitemOpen
  \bibfield  {author} {\bibinfo {author} {\bibfnamefont {F.}~\bibnamefont
  {{\v{S}}imkovic}}, \bibinfo {author} {\bibfnamefont {A.}~\bibnamefont
  {Faessler}}, \bibinfo {author} {\bibfnamefont {H.}~\bibnamefont
  {M{\"u}ther}}, \bibinfo {author} {\bibfnamefont {V.}~\bibnamefont {Rodin}},
  and\ \bibinfo {author} {\bibfnamefont {M.}~\bibnamefont {Stauf}},\ }\bibfield
   {title} {\bibinfo {title} {0 $\nu$ $\beta$ $\beta$-decay nuclear matrix
  elements with self-consistent short-range correlations},\ }\href@noop {}
  {\bibfield  {journal} {\bibinfo  {journal} {Physical Review C}\ }\textbf
  {\bibinfo {volume} {79}},\ \bibinfo {pages} {055501} (\bibinfo {year}
  {2009})}\BibitemShut {NoStop}%
\bibitem [{\citenamefont {Shimizu}\ \emph {et~al.}(2019)\citenamefont
  {Shimizu}, \citenamefont {Mizusaki}, \citenamefont {Utsuno},\ and\
  \citenamefont {Tsunoda}}]{Shimizu:2019xcd}%
  \BibitemOpen
  \bibfield  {author} {\bibinfo {author} {\bibfnamefont {N.}~\bibnamefont
  {Shimizu}}, \bibinfo {author} {\bibfnamefont {T.}~\bibnamefont {Mizusaki}},
  \bibinfo {author} {\bibfnamefont {Y.}~\bibnamefont {Utsuno}}, and\ \bibinfo
  {author} {\bibfnamefont {Y.}~\bibnamefont {Tsunoda}},\ }\bibfield  {title}
  {\bibinfo {title} {{Thick-Restart Block Lanczos Method for Large-Scale
  Shell-Model Calculations}},\ }\href
  {https://doi.org/10.1016/j.cpc.2019.06.011} {\bibfield  {journal} {\bibinfo
  {journal} {Comput. Phys. Commun.}\ }\textbf {\bibinfo {volume} {244}},\
  \bibinfo {pages} {372} (\bibinfo {year} {2019})},\ \Eprint
  {https://arxiv.org/abs/1902.02064} {arXiv:1902.02064 [nucl-th]} \BibitemShut
  {NoStop}%
\bibitem [{\citenamefont {Caurier}\ \emph {et~al.}(2010)\citenamefont
  {Caurier}, \citenamefont {Nowacki}, \citenamefont {Poves},\ and\
  \citenamefont {Sieja}}]{PhysRevC.82.064304}%
  \BibitemOpen
  \bibfield  {author} {\bibinfo {author} {\bibfnamefont {E.}~\bibnamefont
  {Caurier}}, \bibinfo {author} {\bibfnamefont {F.}~\bibnamefont {Nowacki}},
  \bibinfo {author} {\bibfnamefont {A.}~\bibnamefont {Poves}}, and\ \bibinfo
  {author} {\bibfnamefont {K.}~\bibnamefont {Sieja}},\ }\bibfield  {title}
  {\bibinfo {title} {Collectivity in the light xenon isotopes: A shell model
  study},\ }\href {https://doi.org/10.1103/PhysRevC.82.064304} {\bibfield
  {journal} {\bibinfo  {journal} {Phys. Rev. C}\ }\textbf {\bibinfo {volume}
  {82}},\ \bibinfo {pages} {064304} (\bibinfo {year} {2010})}\BibitemShut
  {NoStop}%
\bibitem [{\citenamefont {Men{\'e}ndez}(2017)}]{menendez2017neutrinoless}%
  \BibitemOpen
  \bibfield  {author} {\bibinfo {author} {\bibfnamefont {J.}~\bibnamefont
  {Men{\'e}ndez}},\ }\bibfield  {title} {\bibinfo {title} {Neutrinoless
  $\beta\beta$ decay mediated by the exchange of light and heavy neutrinos: The
  role of nuclear structure correlations},\ }\href@noop {} {\bibfield
  {journal} {\bibinfo  {journal} {Journal of Physics G: Nuclear and Particle
  Physics}\ }\textbf {\bibinfo {volume} {45}},\ \bibinfo {pages} {014003}
  (\bibinfo {year} {2017})}\BibitemShut {NoStop}%
\bibitem [{\citenamefont {\ifmmode~\check{S}\else \v{S}\fi{}imkovic}\ \emph
  {et~al.}(2008)\citenamefont {\ifmmode~\check{S}\else \v{S}\fi{}imkovic},
  \citenamefont {Faessler}, \citenamefont {Rodin}, \citenamefont {Vogel},\ and\
  \citenamefont {Engel}}]{PhysRevC.77.045503}%
  \BibitemOpen
  \bibfield  {author} {\bibinfo {author} {\bibfnamefont {F.}~\bibnamefont
  {\ifmmode~\check{S}\else \v{S}\fi{}imkovic}}, \bibinfo {author}
  {\bibfnamefont {A.}~\bibnamefont {Faessler}}, \bibinfo {author}
  {\bibfnamefont {V.}~\bibnamefont {Rodin}}, \bibinfo {author} {\bibfnamefont
  {P.}~\bibnamefont {Vogel}}, and\ \bibinfo {author} {\bibfnamefont
  {J.}~\bibnamefont {Engel}},\ }\bibfield  {title} {\bibinfo {title} {Anatomy
  of the $0\ensuremath{\nu}\ensuremath{\beta}\ensuremath{\beta}$ nuclear matrix
  elements},\ }\href {https://doi.org/10.1103/PhysRevC.77.045503} {\bibfield
  {journal} {\bibinfo  {journal} {Phys. Rev. C}\ }\textbf {\bibinfo {volume}
  {77}},\ \bibinfo {pages} {045503} (\bibinfo {year} {2008})}\BibitemShut
  {NoStop}%
\bibitem [{\citenamefont {Menendez}\ \emph {et~al.}(2008)\citenamefont
  {Menendez}, \citenamefont {Poves}, \citenamefont {Caurier},\ and\
  \citenamefont {Nowacki}}]{menendez2008disassembling}%
  \BibitemOpen
  \bibfield  {author} {\bibinfo {author} {\bibfnamefont {J.}~\bibnamefont
  {Menendez}}, \bibinfo {author} {\bibfnamefont {A.}~\bibnamefont {Poves}},
  \bibinfo {author} {\bibfnamefont {E.}~\bibnamefont {Caurier}}, and\ \bibinfo
  {author} {\bibfnamefont {F.}~\bibnamefont {Nowacki}},\ }\bibfield  {title}
  {\bibinfo {title} {Disassembling the nuclear matrix elements of the
  neutrinoless double beta decay},\ }\href@noop {} {\bibfield  {journal}
  {\bibinfo  {journal} {arXiv preprint arXiv:0801.3760}\ } (\bibinfo {year}
  {2008})}\BibitemShut {NoStop}%
\end{thebibliography}%
\end{document}